\documentclass[aps,prl,twocolumn,showpacs]{revtex4-1}
\usepackage{amsbsy,latexsym,amsmath,amsthm,amsfonts,mathrsfs,bbold}
\usepackage{amsfonts}
\usepackage{amssymb}
\usepackage{epsfig,graphics,graphicx}

\newcommand{\ket}[1]{\vert#1\rangle}
\newcommand{\bra}[1]{\langle#1\vert}

\newcommand{\expo}[1]{\mathrm{e}^{#1}}

\newcommand{\tr}{\rm{tr}}
\newcommand{\map}[1]{\mathscr{#1}}
\renewcommand{\vec}[1]{\mbox{\boldmath$#1$}}
\newcommand{\vecs}[1]{{\widetilde{\bf #1}}}

\newcommand{\mus}{{\mbox{\boldmath $\scriptstyle \mu$}}}

\newtheorem{theo}{Theorem}
\newtheorem{lemma}{Lemma}

\def\k{\rangle}

\begin{document}
\title{Probabilistic Metrology Attains Macroscopic Cloning of Quantum Clocks}
\author{B. Gendra$^{1}$, J. Calsamiglia$^{1}$, R. Mu\~{n}oz-Tapia$^{1}$, E. Bagan$^{1,2}$ and G. Chiribella$^{3}$}

\affiliation{$^{1}$F\'isica Te\`orica: Informaci\'o i Fen\`omens Qu\`antics, Universitat
Aut\`{o}noma de Barcelona, 08193 Bellaterra (Barcelona), Spain\\ 
$^{2}$Centre for Quantum Technologies, National University of Singapore,
3 Science Drive 2, Singapore 117543, Singapore\\
$^{3}$Center for Quantum Information, Institute for Interdisciplinary Information Sciences, Tsinghua University
Beijing, 100084, China
}

\begin{abstract}
It has been recently shown that probabilistic protocols based on postselection boost the performances of  phase estimation and the replication of quantum clocks.   Here we demonstrate that the improvements in these two tasks have to match exactly in the macroscopic limit where the number of clones grows to infinity, preserving  the equivalence between asymptotic cloning and estimation for arbitrary  values of the success probability.   Remarkably, the cloning fidelity  depends critically on the number of rationally independent eigenvalues of the clock Hamiltonian. We also prove that probabilistic metrology can simulate cloning in the macroscopic limit for arbitrary sets of states, provided  that the performance of the simulation   is measured by testing small groups of clones. 
\end{abstract}

\maketitle


High-resolution measurements and new sensors powered by quantum effects are among the most appealing  gadgets promised by the field of quantum technologies~\cite{qtechno1,qtechno2}.  Consequently, intense effort is being devoted to the design   of prototype setups that achieve quantum-enhanced precision and sensitivity \cite{
qmetro1,
qmetro2}. %
 In recent years,   probabilistic setups based on postselection have attracted a great deal of attention, coming in different variants such as weak value amplification \cite{weak0,weak1,weak2,weak3,weak4,weak5,weak6,weak7} and Probabilistic Metrology (PM)   \cite{fiurasek2,massar-popescu2,gendra,qmet,qmetlong,replication,marek,probably}.   These schemes are based on filters that herald the occurrence of a favourable event, conditional to which the precision is enhanced far beyond the usual limits ---e.~g.~with a scaling upgraded from the Standard Quantum Limit (SQL) to the Heisenberg Limit~(HL).  Typically,  the more dramatic is the improvement, the smaller is the probability of the favourable event, with a  trade-off curve between precision and probability that can be quantified  explicitly in several  interesting cases \cite{gendra,qmet,qmetlong}.  Note that PM maintains its  appeal even in the regime where the probability of favourable events is  small: Indeed, in this regime  one can design the filter so  that, when the  unfavourable event occurs, the state of the system is approximately unchanged, thanks to the so-called Gentle Measurement Lemma \cite{gentle-andreas}.  As a result,  PM offers the experimenter the bonus of hitting the HL from time to time ---and knowing when this favourable event happens--- without compromising optimality of the  average scaling.


The advantages of probabilistic  filters are not limited to metrology.  Instead, they affect a variety of tasks, including  cloning  \cite{duan98,fiuraclon,replication} and amplification \cite{rl,xie,amp1,amp2,amp3,amp4,amp5}.    Very recently, it has been  shown that for quantum clocks the use of a filter can lead to the phenomenon of \emph{super-replication} \cite{replication}, allowing to convert  $n\gg 1$ synchronized clocks  into $m\ll n^2$ replicas, whose joint state appears to be exponentially close  to the ideal target of $m$ perfect copies.  
Achieving such a replication rate with high  fidelity  is impossible without  filtering,  because a deterministic machine that produces more than $O(n)$ nearly perfect replicas   would  lead straight  into a violation of the~SQL. 

Considering the striking difference in performance,  it is natural to ask whether deterministic and probabilistic cloning machines differ in other, more fundamental features. 
The most fundamental  feature of  all  is arguably the \emph{asymptotic equivalence with state estimation} \cite{bruss,bae,chiriclon,tqc2010}, i.~e.~the fact that,  in the macroscopic limit  $m  \to \infty$, the optimal  performance of quantum cloning can be achieved by measuring the input copies and preparing the clones in a state that depends on the measurement outcome. The no-cloning theorem itself \cite{noclon} can be considered as a particular instance of this equivalence:  two states that can~be cloned perfectly by a deterministic machine can also be cloned perfectly in the macroscopic limit and therefore they can be distinguished perfectly by a deterministic estimation strategy, which means that they must be orthogonal to one another.   For deterministic machines, the cloning-estimation equivalence has been proved in full generality when  the   performance of cloning is assessed on small groups of $k\ll m$ clones \cite{chiriclon,tqc2010} and has been recently conjectured to hold even when the $m$ clones are examined collectively \cite{yang,chiriyang}. However, nothing is known in the probabilistic case, where the tradeoff between performance and probability of success  adds  a new twist  to the problem. Here, proving the equivalence requires showing that for every cloning machine there is a protocol based on state estimation that, in the macroscopic limit, achieves the same fidelity with the same probability.   But  is the enhanced precision  of  PM  sufficient to  keep up with  the highly-increased performance of probabilistic cloning machines?     

In this Letter we answer the  question in the affirmative, showing that postselection  does not challenge the fundamental equivalence between cloning and estimation.    We first work out explicitly the example of quantum clocks, where  the performance enhancements are the  most prominent.  We consider clock states $|\psi_t\k  =  e^{-itH}  |\psi_0\k$ generated from an arbitrary input state~$|\psi_0\k$  by time evolution with arbitrary Hamiltonian acting on a $d$-dimensional Hilbert space ${\cal H}$ and we exhibit   PM protocols that achieve the performances of the optimal cloning machine for  every desired value of the success probability.   In this comparison, we use the most restrictive criterion, namely the global fidelity between the clones and~$m$ perfectly synchronized replicas of the original clock.   We evaluate the fidelity explicitly and discover that its value depends critically on the number of rationally independent eigenvalues of the Hamiltonian.   The result is derived using new techniques, based on the Smith  normal form \cite{marcus}, which we expect to be useful for other problems in quantum metrology and optimal quantum information processing.  Furthermore, we analyze the scenario where the performances of cloning are judged from groups of~$k  \ll  m$ clones,  establishing  the equivalence between probabilistic cloning and estimation for arbitrary sets of input states and for arbitrary values of the success probability.  This result extends the validity of equivalence to  all points  of the optimal performance-probability tradeoff.

Let us start from the concrete example of quantum clocks. 
With a suitable choice of basis, the state of a clock at time $t  =  0$ can be written as~$|\psi_0\rangle=\sum_{j=0}^{d-1} \sqrt{p_j} |j\rangle$, where   $H |j\rangle  =  e_j  |j \rangle$  and $p_j$  is the probability that a measurement of energy gives outcome~$e_j$. Without loss of generality, we assume     that all probabilities $\{p_j\}$ are non-zero,  that the eigenvalues $\{e_j\}$ are distinct, and  that~$e_0= 0$.       
In the case of $n$ identical synchronized clocks, we denote the state at time $t$ as $|\Psi^n_t\rangle:  =  |\psi_t\rangle^{\otimes n}$.  
The values of the total energy can be labeled by the partitions of $n$ into $d$ non-negative integers  $(n_0,n_1,\dots , n_{d-1})$.  Denoting by ${\bf n}\in{\mathbb Z}^{(d-1)\times 1}$ [${\bf e}\in{\mathbb R}^{1\times (d-1)}$]  the 1~column integer  (1~row real) matrix $ {\bf n}  =  (n_1,\dots, n_{d-1})^t$ [$ {\bf e}  =  (e_1,\dots, e_{d-1})$] we  express the corresponding energy as~$E_{\mathbf n}:=\sum_{j=1}^{d-1} e_j n_j :={\bf e}\, {\bf n} $.   
The spectrum of the total Hamiltonian will be denoted by~$\mathsf{Sp}_n=\{E_{\bf n}\}_{{\bf n}\in{\mathscr P}_n}$, where${\mathscr P}_n$ is the lattice of vectors~$\bf n$ satisfying  $n_j  \ge 0$ for every~$j $  and $\sum_{j=0}^{d-1}  n_j \le n$. 

%
%

By collecting  the vectors that lead to the same energy~$E$, we define the set ${\mathscr P}^E_n:=\{{\bf n}\!\in\!{\mathscr P}_n\! :\! E_{\bf n}\!=\!E\}$, so that~$|{\mathscr P}^E_n|$ is the degeneracy of $E$.  Then, the state of the~$n$ clocks can be  written as 
\begin{equation}
|\Psi^n_t\rangle\!= \!\!\!\sum_{E\in\mathsf{Sp}_n}\!\!{\rm e}^{-i E t}\sqrt{  p_{E,n}} |E, n\k  ,\quad  {  p_{E,n}} : =\!\! { \sum_{{\bf n}  \in  {\mathscr P}^E_n }    p_{{\bf n},  n } }
,
\label{clock states}
\end{equation}
 where   $p_{{\bf n},n}$ is the multinomial distribution $p_{{\bf n},n}   :=  {n!\prod_{j=0}^{d-1} p_j^{n_j}/n_j!}$ and  
\begin{equation}
|E, n\k:={1\over {\sqrt {p_{E,n}}}}\sum_{{\bf n}\in{\mathscr P}_n^E} \sqrt {p_{{\bf n}, n}} |{\bf n},  n\k \, .
\label{H^n eigen}
\end{equation}

In the metrology scenario, the goal is to estimate the time $t$ as accurately as possible. 
The estimate, denoted by $\hat t$, can be used to produce $m$ approximate clones, by  preparing a state~$|\hat\Psi_{\hat t}^m\k =   \sum_{E  \in  {\sf Sp}_m }   e^{-iE\hat  t}  \, \sqrt{  \hat p_{E,m}}  |E,m\k $, so that, averaging over all possible values of $\hat t$, the output state resembles~$|\Psi^m_t\k$.   As a performance measure, we adopt the worst-case fidelity 
\begin{equation}
\label{figure} 
F   =  \inf_{t\in  {\mathbb R}}  \int  {\rm d} \hat t  \,    p(\hat t  |  t)   \,  \left|  \langle   \hat\Psi_{\hat t}^m  |   \Psi_t^m  \rangle \right|^2 \, ,
\end{equation} 
where $p(\hat t|t)$ is the probability of estimating $\hat t$ when the true value is $t$.  
For covariant families of states, such as the quantum clocks under consideration, the worst-case fidelity is  equal  to the average of the fidelity with respect to the uniform prior. 
In the case of PM  the estimation strategy does not provide an estimate all the times, but sometimes declares ``failure", in which case one abstains from producing copies. Hence, $p(\hat t |t)$ has to be understood as $p(  \hat t  |   t,  {\rm succ})$, the probability of estimating $\hat t$ conditional to the fact that the strategy succeeded in producing an estimate.   To take this into account,  one can equivalently replace the input state $|\Psi_t^n\k$ by a state the form $\Pi^{1/2}  | \Psi_t^n\k/  \| | \Pi^{1/2}  | \Psi_t^n\k \|$, where $\Pi $ is  suitable operator satisfying $  0  \le  \Pi  \le \openone^{\otimes n}$.  By the symmetry of the figure of merit in  Eq. (\ref{figure}),  it is easy to see that $\Pi $ can be chosen to be diagonal in the energy basis, namely $\Pi=\sum_{E\in {\sf Sp}_n} \pi_E |E,n\k  \kern-.1em\langle E,n|$.  With this choice the probability of success  is independent of $t$ and is given by  $ P_{\rm succ}:=  \langle \Psi^n_t|\Pi|  \Psi^n_t\rangle$.   Fixing the coefficients of the guess state and of the filter, we show that the supremum of the fidelity over all quantum measurements  is \cite{epaps}
\begin{equation}
F=\sum_{\cal E}\left(\sum_{E\in{\sf Sp}_n}
\kern-.5em\raisebox{.2em}{$'$} \sqrt{  p_{E,n} \,   p_{E+{\cal E},m} \,   \hat p_{E+{\cal E},m} \, { \pi_E\over P_{\rm succ}}}
\,\right)^2,
\label{F_M&P}
\end{equation}
where the outer sum runs over the set $ \{  {\cal E}  =  E_m - E_n  ~|~  E_m\in {\sf Sp}_m,\   E_n\in {\sf Sp}_n      \} $ and the prime~(${}'$) means that the sum is restricted to  those  $E\in {\sf Sp}_n$ such that $E+{\cal E}\in {\sf Sp}_m$.     For a fixed filter $\Pi$,   we denote by $F^\Pi_{\rm PM} $ the supremum of the fidelity in Eq. (\ref{F_M&P}) over all possible guess states. 
The question now is whether, suitably choosing the filter $\Pi$, the value of $F^\Pi_{\rm PM}$ can reach the fidelity of the optimal quantum copy machine in the macroscopic limit~$m\to \infty$. 

%
%
%
%

Let us examine now the cloning scenario. Here,  one can produce copies coherently using a general quantum operation (completely positive trace non-increasing map) ${\mathscr C}_{n,m}$ that maps  states on ${\cal H}^{\otimes n}$ to states on ${\cal H}^{\otimes m}$.   
The quantum operation $\mathscr C_{n,m}$  can always be written as a composition of a probabilistic filter~$\Pi$ followed by a trace preserving map~${\mathscr D}_{n,m}$ \cite{replication}. Again, the symmetry of the figure of merit allows one to choose without loss of generality a filter of the form $\Pi =\sum_E \pi_E |E,n\k  \kern-.1em\langle E,n|$.   
Our first step is to upper bound  the maximum fidelity achievable for a fixed filter, denoted by $F^\Pi_{\rm CL} $, as \cite{epaps}
\begin{equation}
F^\Pi_{\rm CL} \le \max_{E\in{\sf Sp}_m}\!\!\{p_{E,m}\}\left(\sum_{E\in{\sf Sp}_n}\sqrt{  p_{E,n} {\pi_E\over P_{\rm succ}}}
\right)^2.
\label{upper bound}
\end{equation}
%

%
The next step is to  show  that the fidelity $F^\Pi_{\rm PM}$ achieves the upper bound in the limit of large $m$.   
For the sake of illustration, we start from the simple case where  all  the energies $\{e_j\}$  are  commensurable,  i.e.,~$e_j=k_j \, \varepsilon$ where~$k_j$ is an integer and $ \varepsilon$ is a fixed unit of energy.    Then, the energy  eigenvalues for $m$ clocks can be written as  $E_{\bf m}= \varepsilon\;  {\bf k}\,{\bf m} $. 
For sufficiently large $m$, it is easy to see that the minimum spacing between two consecutive energies is given by $\Delta E^*  =     \varepsilon \gcd \{  k_\alpha\} $, with  $\gcd$ denoting the greatest common divisor. This fact is an immediate consequence of Bezout's identity in number theory~\cite{bezout}.      Now, suppose that $m$ is asymptotically large.  In this limit, the multinomial distribution $p_{{\bf m},m}$ approaches the multivariate normal distribution~${\mathscr N}(m {\bf p},\vec\Sigma)$, where ${\bf p}:=(p_1,\dots p_{d-1})^t$ and the covariance matrix $\vec\Sigma$ has entries $\Sigma_{jj}=mp_j(1-p_j)$, $\Sigma_{jl}=-m p_j p_l$, $j\not=l$. Clearly, this implies that the probability distribution $p_{E,m}$ in Eq. (\ref{clock states}) is concentrated in a window of size~$O(  \sqrt m)$ centred around the mean  value~$\langle H\rangle$.  Within this window, every two consecutive energies differ by the minimum amount $\Delta E^*$~\cite{nota}.  Finally, note that by dimensional arguments the degeneracy of a typical energy grows as~$|{\mathscr P}^E_m|\sim m^{d-2}$. 
Thus, the sum  over ${\bf m}\in {\mathscr P}^E_m$ that defines $p_{E,m}$ in Eq.~(\ref{clock states}) can be approximated by the  integral of~${\mathscr N}(m{\bf p},\vec\Sigma)$ over a $(d-2)$-dimensional domain.  As a result,~$p_{E,m}$ is approximated by the discrete Gaussian distribution 
\begin{equation}
p_{E,m}\approx \Delta E^*{{\rm e}^{-{(E  - m \langle  H \rangle )^2\over 2 m   {\rm Var}  (H)}}\over\sqrt{2\pi m  {\rm Var} (H) }},
\quad \Delta E^*   = \varepsilon \gcd\{k_\alpha \}
\label{p^m_E},
\end{equation}
%
where ${\rm Var}  (H)  =  \langle H^2  \rangle  -  \langle H \rangle^2$ is the variance of $H$. 

Thanks to Eq. (\ref{p^m_E}) we are now in position to show that $F^\Pi_{\rm PM}$ approaches $F^\Pi_{\rm CL}$ in the macroscopic limit. 
Indeed, since for $m\gg n$ the probability $ p_{E,m}$  is almost constant over every interval of size $O(n)$,   we can pull out a factor~$\max_{E\in{\sf Sp}_m} \{{  p_{E,m}}\}$ from both sums in~Eq.~(\ref{F_M&P}) introducing an error that vanishes in the asymptotic limit.  
Moreover, we can  choose a guess state with  probabilities~$\hat p_{E,m}$ given by a discrete Gaussian with width  growing as~$\sqrt{m^{1-\eta}}$,~$0<\eta<1$.  
Since the width is much larger than $O(n)$,  we have  \mbox{$\hat p_{E  +{\cal E}, m}\approx \hat p_{{\cal E},m}$} for every $E\in{\sf Sp}_n$ and we can pull out the term $\hat p_{{\cal E},m}$ from the sum over~${\sf Sp}_n$. Since the width is much smaller than $\sqrt m$, we have ~$\sum_{\cal E} p_{{\cal E},m}  \approx 1$. Hence, our choice of guess state attains the upper bound in~Eq.~(\ref{upper bound}) and we have \cite{epaps}
 \begin{equation}\label{F^Pi_PM = F^Pi_CL}
  F^\Pi_{\rm PM} \!\approx\! F^\Pi_{\rm CL}  \!    \approx \! { \Delta E^*\over\sqrt{2\pi m {\rm Var}(H)}}   \left(\sum_{E\in{\sf Sp}_n}\!\!\sqrt{  p_{E,n} {\pi_E\over P_{\rm succ}}}
\right)^2\!\!.
 \end{equation} 

 The reader should not be misled by the simplicity of Eq. (\ref{p^m_E}), which superficially may seem an application of the Central Limit Theorem (CLT). The CLT gives an approximation of the \emph{cumulative} distribution of $p_{E,m}$, not of the probability mass itself.  In fact, those who believe that $p_{E,m}$ should converge to a Gaussian are in for a surprise in the case where the eigenvalues of $H$ are not commensurable.  In this case, the eigenvalues~$\{e_j\}$  can be expressed as  integer linear combinations   of  a minimal  number~$r$ of rationally independent units of energy~$\{\varepsilon_l\}_{l=1}^r$.  Rational independence means that,  for every set of integer coefficients $\{c_l\}$, the relation $\sum_l  c_l \varepsilon_l=  0 $ implies~$c_l=0$ for every $l$.  In terms of the units $\{\varepsilon_l\}$,   we expand each eigenvalue as~$e_j=\sum_l    \varepsilon_l    k_{lj}  $ where $\{k_{l j}\}$ are integer coefficients, uniquely defined thanks to the rational independence of the units. 
Using the decomposition, we express the energy of $m$ clocks as  $E_{\bf m} =  \boldsymbol{\varepsilon} \, \widetilde {\bf m}$ where  $\boldsymbol\varepsilon=(\varepsilon_1,\dots,\varepsilon_r)$, and $\widetilde{\bf m}=(\widetilde m_1,\dots,\widetilde m_r)^t$ is the 1~column matrix  with components $\widetilde m_l   =   \sum_j   k_{lj}   m_j$, or, more compactly, $\widetilde{\bf m}  ={\bf K}\, {\bf m}$ where~$\bf K$ is  the $r\times(d-1)$ matrix with entries~$\{k_{lj}\}$.  
The matrix $\bf K$ maps the lattice  $\mathscr  P_m  \subset \mathbb Z^{d-1}$  into a  new lattice  $\widetilde{\mathscr P}_m\subset \mathbb Z^r$ with the special feature that the points in~$\widetilde{\mathscr P}_m$ are into one-to-one correspondence with the  energies in ${\sf Sp}_m$ (again, due to the rational independence of the units).   
Hence, instead of the probability distribution $p_{E,m}$ we can consider the probability distribution $ p_{\widetilde {\bf m}, m} :  =     \sum_{{\bf m}:   \bf K  m  =  \widetilde m  } p_{{\bf m}, m}$. 
Now, for large $m$ this probability distribution is concentrated  in a volume of size~$O(m^{r/2})$ centred around the mean $m  {\bf K}\, {\bf p}$. 
The typical vectors   ${\bf m}  \in \mathscr P_m$ associated to points in this volume form a regular Bravais lattice and the number of vectors associated to a point $\widetilde {\bf m}$ grows as $m^{d-r-1}$  \cite{epaps}.  By the same arguments as in the paragraph after Eq.~(\ref{upper bound}), we can approximate the sum in the expression of $ p_{\widetilde {\bf m},m}$ with an integral, thus obtaining \cite{epaps}
\begin{equation}\label{pmm}
 p_{\widetilde{\bf m},m}\!\approx\! \Delta V^*
 {   \exp\!\left\{ \! -{ (\widetilde{\bf m}-\widetilde{\bf m}_0)^t\widetilde{\vec\Sigma}^{-1}\!(\widetilde{\bf m}-\widetilde{\bf m}_0)\over2   }   \!\right\}  \over (2\pi)^{r/2}\sqrt{\det \widetilde{\vec\Sigma}}},
\end{equation}
%
where $\widetilde{\bf m}_0=m{\bf K}{\bf p}$, $\widetilde{\vec\Sigma}={\bf K}{\vec\Sigma}{\bf K}^{t}$ and $\Delta V^*$ is the volume of a minimal cell of the lattice $\widetilde{\mathscr P}_m$. Using the Smith normal form of ${\bf K}$ one can show that $\Delta V^*=\gcd \{[{\bf K}]_r\} $, where $\{[{\bf K}]_r\}$ is the set of all minors of $\bf K$ of order~$r$~\cite{epaps}. Eq. (\ref{pmm})  has two major consequences:  First, the probability mass $p_{E,m}$ \emph{does not converge to a Gaussian}, as one would naively expect from a misapplication of the~CLT.  Instead, it converges to the non-continuous function $p_{ \widetilde{ \bf m}  (E), m}$, where $\widetilde{ \bf m}  (E)$ is the value of $\widetilde{ \bf m}$  such that $E  =  E_{\widetilde{ \bf m}}$.  Second, all the steps of the proof for commensurable energies can now be reproduced by  replacing the energy $E\in {\sf Sp}_n$ with the corresponding vector $\widetilde  {\bf n} (E)$.  In this way  we obtain $F^\Pi_{\rm PM}\approx F^\Pi_{\rm CL}\approx F^\Pi$, with 
\begin{equation}\label{F^Pi}
F^\Pi  : ={\Delta  V^*\over\sqrt{  (2\pi)^r \det\widetilde{\vec\Sigma}}}
\left(\sum_{\widetilde{\bf n}\in{\widetilde{\mathscr P}}_n}\!\!\! \sqrt{   p_{\widetilde {\bf n},n} { \pi_{\widetilde {\bf n}}\over  P_{\rm succ}}}\right)^2 .
\end{equation}
Note that, since  $\det \widetilde{\vec\Sigma}$ scales as $m$, the fidelity scales as $m^{-r/2}$.  Hence, the asymptotic behaviour  depends dramatically on the number of rationally independent units. 
Quite remarkably, this means that the fidelity is not continuos in the Hamiltonian: Although one can approximate arbitrarily well the  Hamiltonian   $H$ with another Hamiltonian $H'$ that has commensurable energies, the corresponding fidelities are not going to be close. 
The origin of the discontinuity is that the value of the fidelity depends on the \emph{closure} of the set of clock states, due to the infimum in Eq. (\ref{figure}).    
When the energy eigenvalues are  commensurable, the time evolution is periodic and the orbit $\{|\Psi^n_t\rangle, \, t\in\mathbb R\}$ is a close one-dimensional curve.  But when the energies are combinations of $r$ rationally independent units, the orbit is dense in an $r$-dimensional submanifold of the manifold of pure states.  This phenomenon is the quantum analog of     
 a classic feature of integrable Hamiltonian systems \cite{katok}, where rationally independent frequencies lead to ergodic time evolutions in phase space.  Here the fidelity is discontinuous because it depends on the long-time behaviour of the time evolution, during which the quantum clock can probe a higher dimensional manifold. Note that for finite times the differences between both families of clock-states generated  by $H$ and $H'$ can be arbitrary small and that continuity  is retrieved if we restrict the infimum in Eq. (\ref{figure}) to a fixed  time interval $[T_1,T_2]$.

Having proven the asymptotic equivalence  between probabilistic metrology and cloning of quantum clocks, we now  give closed expressions for optimal fidelity, maximized over all possible filters. We focus on the large $n$ limit under the condition $n\ll \sqrt m$ and   consider  two relevant regimes:  First, we allow arbitrarily low probability of success, showing that the ultimate fidelity is \cite{epaps,nota2}
\begin{equation}\label{low P fide}
F=\left[(2\pi)^{r}\det\widetilde{\vec\Sigma}\right]^{-1/2} \! |\widetilde{\mathscr P}_n|\,\Delta V^*,
\end{equation}
where  $|\widetilde{\mathscr P}_n|$ is the number of sites in the lattice ${\widetilde{\mathscr P}}_n$. Since $|\widetilde{\mathscr P}_n|$ scales as $n^{r}$, the fidelity scales as $F\sim (n/{\sqrt{m}})^r$. Second, we consider the case where the probability of success is high, i.e. $ P_{\rm succ} = 1-\eta$ for some small $\eta$. Here the optimal fidelity acquires the particularly simple form 
 \begin{equation}
 F=\left(4 {n\over m}\right)^{r/2}[1+\eta(1-2^{-r/2})]+O(\eta^2) \, .
 \label{high P fide}
\end{equation}
Quite surprisingly, $F$  does not depend on the coefficients~$\{p_j\}$ of the input state, but only on the number of rationally independent units $r$. 
 
We conclude by discussing the equivalence between probabilistic metrology and cloning for  arbitrary sets of states.   Here we assess the performance of cloning by looking at a random subset of $k \ll  m$ clones, evaluating the global fidelity between the state of the $k$ clones and the state of $k$ ideal copies.
  Clearly, since the $k$ clones are picked at random, one can assume that the optimal cloner is invariant under permutation of the $m$ output systems. Technically, this means that the cloner is described by a quantum operation $\mathscr C_{n,m}$ such that, for every permutation $\pi$, one has  $\mathscr U_\pi \mathscr C_{n,m}  =  \mathscr C_{n,m}$, where $\mathscr U_\pi $ is the permutation map defined by $\mathscr U_\pi   (\rho_1\otimes \rho_2\otimes  \cdots\otimes \rho_m)   = \rho_{\pi(1)}  \otimes \rho_{\pi (2)} \otimes\dots\otimes \rho_{\pi(m)}$.  Using a de Finetti-type argument, we prove the following result \cite{epaps}:
\begin{theo}\label{theo:one}
For every quantum operation $\mathscr{C}_m$ with input in $\mathcal H_{\rm in}$ and permutationally invariant output in  $\mathscr H^{\otimes m}$ there exists a PM protocol, described by a quantum operation   $\widetilde{\mathscr C}_m$, such that \emph{i)}    $\mathscr{C}_m$ and $\widetilde {\mathscr{C}}_m$ have the same success probability  and \emph{ii)} the error probability in distinguishing  between $\mathscr{C}_m$ and~$\widetilde{\mathscr{C}}_m$ by inputting a state $\rho$ and measuring $k $ output  systems is lower bounded by 
$ p_{\rm err}   \le \frac 1  2   +  (kd^2)/[   2m  P_{{\rm succ}}  (\rho) ]$, where $P_{\rm succ}(\rho)  := {\rm tr}  [ \mathscr{C}_m  (\rho)]$.  
 \end{theo}   
In the case of cloning, the result implies that the  $k$-copy fidelity of an arbitrary  $n$-to-$m$ cloner on a generic input state $|\psi_x\rangle^{\otimes n}$ can be achieved by  PM, up to an error of size   $k/[   m  P_{\rm succ}  (|\psi_x\rangle\langle \psi_x|^{\otimes n})] $ \cite{epaps}.  
Hence, the error  vanishes as $k/m$ for every process with success  probability   larger than a given finite value for every possible input.      This result extends the equivalence between cloning and estimation to every point of the tradeoff curve between fidelity and success probability.   
The result holds also in the Bayesian scenario where the input  state $|\psi_x\rangle^{\otimes n}$ is given with prior probability $p_x$ and one considers average fidelities and average success probabilities. Furthermore,   we show that the average success probability of the optimal  $n$-to-$m$ cloner is lower bounded by a finite value independent of $m$ and therefore the best asymptotic cloner can be simulated by  PM  \cite{epaps}. 

In conclusion, we proved that probabilistic protocols empowered by postselection do not challenge the fundamental equivalence between cloning and estimation. We worked out explicitly the case of quantum clocks,  where the performance enhancements for both tasks are most dramatic, and developed a technique to evaluate the optimal asymptotic fidelity.   We found out  that the asymptotic fidelity depends critically on the number of rationally independent units generating the spectrum of the Hamiltonian, due to an effect that is analog to ergodicity of classical dynamical systems. Finally, we discussed the case of arbitrary families of states, establishing an equivalence between  probabilistic metrology and cloning when the performance is quantified by the fidelity of $k\ll m$ randomly chosen clones.

We thank Elio Ronco-Bonveh\'i for his collaboration at early stages of this work. We acknowledge support  by  the European Regional Development Fund (ERDF), the National Basic Research Program of China (973) 2011CBA00300 (2011CBA00301), the Spanish MICINN (project FIS2008-01236)  with FEDER funds,  the Generalitat de Catalunya CIRIT (project  2009SGR-0985), the National Natural Science Foundation of China through Grants  11350110207, 61033001,  and 61061130540, and  by the Foundational Questions Institute through the large grant ``The fundamental principles of information dynamics".  GC is supported by the 1000 Youth Fellowship Program of China.

\appendix

\pagebreak

\section{SUPPLEMENTAL MATERIAL}

\subsection{Fidelity of probabilistic metrology}

In this section, we derive Eq.~(\ref{F_M&P}) for the PM fidelity.   Since later we will show that asymptotically the r.h.s. of Eq.~(\ref{F_M&P}) achieves the optimal cloning fidelity, here it is enough to show that the r.h.s. of Eq.~(\ref{F_M&P})  can be  achieved by some suitable measurement.  

For every $\sigma  >0$, consider the   operators~$\Phi^n_{\hat t,\sigma}=|\Phi^n_{\hat t,\sigma}\rangle\langle\Phi^n_{\hat t,\sigma}|$, where
%
\begin{equation}
|\Phi^n_{\hat t,\sigma}\rangle:= \sqrt{p_
\sigma(\hat t\,)}\; \sum_{E\in{\sf Sp}_n}{\rm e}^{-iE\hat t}|E,n\rangle \, ,
\end{equation}
and $p_\sigma(t)$ is a suitable probability distribution.  The latter is chosen as follows:  
For commensurable energies, when the evolution is periodic, $p_\sigma (t)$ is the  uniform distribution $p_\sigma(t)=1/T$ over the period $T$ [$p_\sigma(t)=0$ for $t<0$ and $T<t$].   For incommensurable energies, $p_\sigma (t)$ is the Gaussian distribution $p_\sigma(t)=(2\pi\sigma^2)^{-1/2}\exp\{-t^2/(2\sigma^2)\}$.   Now, for commensurable energies, the operators $\{\Phi^n_{\hat t,\sigma}\}_{\hat t\in{\mathbb R}}$ define a quantum measurement: indeed, it is immediate to check the normalization condition $\int {\rm d} \hat t  \,  \Phi^n_{\hat t,\sigma}   =  \openone_n$,
where~$\openone_n$ denotes the identity on the subspace containing the state of the $n$ input copies. 
For incommensurable energies, the operators~$\{\Phi^n_{\hat t,\sigma}\}_{\hat t\in\mathbb R}$ form an  ``approximate measurement", satisfying the  approximate normalization condition  $\int {\rm d} \hat t  \,  \Phi^n_{\hat t,\sigma}   =  \openone_n   +  O(  e^{-\sigma^2 })$. Note that in the limit~$\sigma  \to \infty$ the approximate measurement becomes arbitrarily close to a legitimate measurement, as the  normalization defect disappears in such limit. 

Let us denote by $F_\sigma$ the value  obtained by inserting the approximate measurement $\{\Phi^n_{\hat t,\sigma}\}_{\hat t\in\mathbb R}$ in Eq. (\ref{figure}).   
Since  our set of approximate measurements becomes closer and closer to the set of allowed measurements as $\sigma\to\infty$, the limit value $F_*  :=  \lim_{\sigma\to\infty} F_\sigma$ is an achievable value of the fidelity.  We now show that $F_*$ is equal to the r.h.s. of Eq.~(\ref{F_M&P}): recalling the expansion of $|\Psi^n_t\rangle$ in Eq.~(\ref{clock states}) and the definition of $|\hat\Psi^n_{\hat t}\rangle$ after Eq.~(\ref{H^n eigen}), we obtain
\begin{align}
F_\sigma    =   \inf_t\int {\rm d} \hat t  \,    p_\sigma(\hat t\,)    \left|\sum_{\cal E}   {\rm e}^{-i{\cal E}(t-\hat t)}  f_{\cal E} \right|^2  ,
\end{align}
with 
\begin{align}
  f_{\cal E}  :  = \sum_{E\in{\sf Sp}_n} \, \sqrt{{\pi_Ep_{E,n}p_{E+{\cal E},m}\hat p_{E+{\cal E},m}\over P_{\rm succ}}} \, .
\end{align}
%
%
Integrating over $\hat t$ we then obtain 
\begin{align*}
F_*    & =\lim_{\sigma\to\infty}   \inf_t  \sum_{\cal E,\cal E'}   \,     {\rm e}^{-\frac{\sigma^2({\cal E}-{\cal E}')^2}2} \, {\rm e}^{-it(\cal E-\cal E')}   \, f_{\cal E}   f_{\cal E' }  \\
&  = \sum_{\cal E}  f_{\cal E}^2\, ,   
\end{align*} 
which coincides with the r.h.s. of Eq.~(\ref{F_M&P}).  

\subsection{Cloning fidelity and its upper bound}

In this section we derive the upper bound~(\ref{upper bound}) to the probabilistic cloning fidelity~\cite{replicationS}. For the sake of self-completeness, we also derive the fidelity itself. We start from the definition of the worst case cloning fidelity:
\begin{equation}\label{probfid}
F= 
\inf_{t\in  {\mathbb R}} { \langle   \Psi_{t}^m|{\mathscr C}_{n,m}(\Psi^n_t) |  \Psi_t^m  \rangle\over P_{\rm succ}}
\end{equation}
where in full generality the quantum operation ${\mathscr C}_{n,m}$ 
[see paragraph before Eq.~(\ref{upper bound})]
has been decomposed as a probabilistic filter followed by a deterministic (trace preserving) map, i.e., as ${\mathscr C}_{n,m}={\mathscr D}_{n,m}\circ\Pi$ \cite{replicationS}, and the probability of success is~$P_{\rm succ}=\langle\Psi_t^n| \Pi|\Psi^n_t\rangle$. The covariance of quantum clocks enables us to drop the infimum in the last equation and consider without loss of generality only invariant filters of the form $\Pi=\sum_{E\in{\sf Sp}_n}\!\! \pi_E|E,n\rangle\langle E,n|$, as well as covariant maps, which satisfy 
\begin{equation}
{\mathscr D}_{n,m}(U^n_t\ \cdot\ U^{n\dagger}_t)=U^m_t{\mathscr D}_{n,m}(\ \cdot\ )U^{m\dagger}_t ,
\label{cov map}
\end{equation}
where $U^n_t=\sum_{E\in{\sf Sp}_n}\!\!{\rm e}^{-i E t}|E,n\rangle\langle E,n|$ is the time evolution operator. Using the Choi-Jamiolkowski isomorphism, Eq.~(\ref{cov map}) is equivalent to
\begin{equation}
{\mathcal D}=U^m_t\otimes U^{n\,*}_t\;{\mathcal D}\; \left(U^{m}_t\otimes U^{n\,*}_t\right)^\dagger,
\label{cov C-J}
\end{equation}
which in turn implies the direct sum decomposition ${\mathcal D}\!=\!\sum_{\cal E} {\mathcal D}_{\cal E}$, where
\begin{equation}
{\mathcal D}_{\cal E}\!\!:=\!\!\!\!\!\sum_{E,E'\in {\sf Sp}_n}\kern-1.1em\raisebox{.2em}{$'$}d^{\cal E}_{E,E'}
|E+{\cal E},m\rangle\langle E'+{\cal E},m| \otimes |E,n\rangle\langle E',n|
\end{equation}
and the `primed' sum include only those terms for which $E+{\cal E}, E'+{\cal E}\in{\sf Sp}_m$. Taking the above into account, the probabilistic cloning fidelity can be cast as
\begin{eqnarray}
F\!&=&\!{1\over P_{\rm succ}}\sum_{{\cal E}}\!\!
\sum_{E, E'\in {\sf Sp}_n}\kern-1.1em\raisebox{.2em}{$'$}\;
d^{\cal E}_{E,E'}\nonumber \\
&\times&
\sqrt{\pi_E p_{E,n}p_{E+{\cal E},m}}\sqrt{\pi_{E'} p_{E',n}p_{E'+{\cal E},m}}\,.
\label{prob clon fid}
\end{eqnarray}

An upper bound to the fidelity~(\ref{prob clon fid}) can be obtained by pulling the maximum value of $p_{E,m}$ out of the sums.
Recalling the positivity of the Choi-Jamiolkowski operator, ${\cal D}\ge0$, one has ${\cal D}_{\cal E}\ge0$ and $|d^{\cal E}_{E,E'}|^2\le d^{\cal E}_{E,E}d^{\cal E}_{E',E'}$. Thus,
\begin{eqnarray}
 F &\le& \max_{E\in{\sf Sp}_m}\!\!\left\{p_{E,m}\right\}
\nonumber \\
&\times&\kern-.5em
\sum_{E, E'\in {\sf Sp}_n}\!\!
\sum_{{\cal E}}\kern-0.1em\raisebox{.2em}{$'$}\;
\sqrt{\!{\pi_{\!E} d^{\cal E}_{E,E}p_{E,n}\over P_{\rm succ}}
}
\sqrt{\!{\pi_{\!E'} d^{\cal E}_{E'\!,E'}p_{E'\!,n}\over P_{\rm succ}}
},
\label{bound F_CL}
\end{eqnarray}
where we have interchanged the order of summation.  
We can now use the Schwarz inequality to write 
\begin{equation}
\sum_{{\cal E}}\kern-0.1em\raisebox{.2em}{$'$} \sqrt{d^{\cal E}_{E,E}}
 \sqrt{d^{\cal E}_{E',E'}}\le\left(\sum_{{\cal E}}\kern-0.1em\raisebox{.2em}{$'$}d^{\cal E}_{E,E} \right)
 \left(\sum_{{\cal E}}\kern-0.1em\raisebox{.2em}{$'$}d^{\cal E}_{E',E'} \right).
\end{equation}
The first (second) sum on the right hand side of the last equation can be extended to values of $\cal E$ for which $E+{\cal E}\in{\sf Sp}_m$  ($E'+{\cal E}\in{\sf Sp}_m$), as $d^{\cal E}_{E,E}\ge0$ ($d^{\cal E}_{E',E'}\ge0$). Then, each of these sums is unity since ${\mathscr D}_{n,m}$ is trace preserving. Substituting in~(\ref{bound F_CL}), we obtain the bound in Eq.~(\ref{upper bound}).

\subsection{Geometry of the problem and Smith variables}

\subsubsection{Rationally independent energy units and lattices}

We recall that as a set of points 
the partitions~$\{(n_0,{\bf n})\}$, where
${\bf n}\in{\mathscr P}_n$, is a regular lattice on the simplex~$\Delta^{d-1}_n=\{(x_0,\dots,x_{d-1})\in{\mathbb R}^{d}:  x_j\ge0, j=0,\dots,d-1\,\mbox{and}\,\sum_{j=0}^{d-1} x_j=n\}$, of edge length $n$. Each site of this lattice is of the form~$(n-\sum_{j=1}^{d-1} n_j,{\bf n})$. The vectors (1~column integer matrices) ${\bf n}\in{\mathscr P}_n$ form themselves also a regular lattice, defined by the inequalities~$0\le n_j\le n-\sum_{l=1}^{j-1} n_l$, $j=1,\dots, d-1$, inside the corner of a $(d-1)$-dimensional cube of side length~$n$: $\Delta^{d-1}_{c,n}=\{(x_1,\dots, x_{d-1})\in{\mathbb R}^{d-1}: 0\le x_j\le n-\sum_{l=1}^{j-1} x_l, j=1,\dots,d-1 \}$. 
Note that if~$n\le m$, one has the inclusion~\mbox{${\mathscr P}_n\subset{\mathscr P}_m$}.  

Recall also that the spectrum of $H^{\otimes n}$,  is given by~${\sf Sp}_n=\{E_{\bf n}={\bf e}\, {\bf n} : {\bf n}\in{\mathscr P}_n\}$.  All vectors  {\bf n} that give rise to a particular value $E$ of the energy, i.e., those in the set~${\mathscr P}^E_n=\{{\bf n}\in{\mathscr P}_n : E_{\bf n}=E\}$,   necessary lie in the affine hyperplane obtained by translating the hyperplane orthogonal to  ${\bf e}$. Since ${\mathscr P}_n$ is a regular lattice, it is not clear a priori how many of its sites fall on the hyperplane defined by ${\bf e}$; the actual degeneracy of~$E$ strongly depends on the commensurability of the energies $\{e_j\}$ in the spectrum of $H$. To identify all distinct values of $E_{\bf n}\in{\sf Sp}_n$ we use the fact that the energies of the Hamiltonian $H$ can always be written as a linear combination  with integer coefficients of a minimal set of rationally independent `energy units'~$\{\varepsilon_l\}_{l=1}^r$, $r\leq d-1$,  namely \begin{equation}\label{minimal}
e_j=\sum_l k_{lj}\varepsilon_l  \qquad    k_{lj}\in{\mathbb Z} \,.
\end{equation}  
 By \emph{minimal} we mean that no subset of  $\{\varepsilon_l\}_{l=1}^r$ is sufficient to write every energy in the spectrum of $H$ as in Eq. (\ref{minimal}).  Note that the rational independence of $\varepsilon_l$  implies that $k_{lj}$ is fixed once the choice of energy units is made.
With this, the energies of~${\sf Sp}_n$ can be written as $E_{\bf n}=\sum_{l=1}^r\varepsilon_l \widetilde{n}_l$, where~$\widetilde n_l=\sum_{j=1}^{d-1} k_{lj}n_j$. It is then useful to introduce the vector notation $\widetilde{\bf n}={\bf K}\,{\bf n}$, where~${\bf K}$ is the $r\times(d-1)$ matrix whose integer entries are~$k_{lj}$, and~$E_{\bf n}=\boldsymbol{\varepsilon}\,\widetilde{\bf n}$, where $\boldsymbol\varepsilon=(\varepsilon_1,\dots,\varepsilon_r)$. Because of the rational independency of the energy units, we conclude that there is a bijection between the distinct energies in~${\sf Sp}_n$ and the points in the set~$\widetilde{\mathscr P}_n=\{\vecs{n}={\bf K}\,{\bf n} : {\bf n}\in{\mathscr P}_n\}$.


%

We view each column \mbox{${\bf K}_j=(k_{1j},\dots,k_{rj})^t\in{\mathbb Z}^{r}$} 
of~$\bf K$ as a set of vectors that span the (infinite) Bravais lattice
\begin{equation}
\widetilde{\mathscr P}_\infty=\mbox{$
\left\{\vecs{n}=\sum_{j=1}^d \,n_j\,{\bf K}_j,\; n_j\in{\mathbb Z}\right\}
$}.
\label{eq:brav1}
\end{equation}
%
We have the obvious inclusion $\widetilde{\mathscr P}_n\subset\widetilde{\mathscr P}_\infty$.
 For finite~$n$, $\widetilde{\mathscr P}_n$ departs from the Bravais lattice $\widetilde{\mathscr P}_\infty$ in two ways: i)~the lattice~$\widetilde{\mathscr P}_n$,  defined by the linear transformation~$\bf K$ acting on ${\mathscr P}_n$, inherits its boundaries and hence lies inside  the convex $r$-dimensional polytope $\widetilde\Delta^r_{n}:={\bf K}\Delta^{d-1}_{c,n}$; ii)~near the boundaries of $\widetilde\Delta^r_{n}$, some points of $\widetilde{\mathscr P}_\infty$ are missing in $\widetilde{\mathscr P}_n$ (since none of the corresponding inverse images satisfy the constrains that define ${\mathscr P}_n$, given in the first paragraph of this section), so we typically have  $\widetilde{\mathscr P}_n\varsubsetneq\widetilde{\mathscr P}_\infty\cap\widetilde\Delta^r_n$. These boundary related issues have a minor effect in our analysis for asymptotically large $n$, as we argue below.

\subsubsection{Smith vectors/variables. Volume of primitive cell}

For~$r<d-1$,  the vectors ${\bf K}_j$ are not linearly independent and, therefore,  they cannot be  a minimal set of primitive vectors of the lattice $\widetilde{\mathscr P}_\infty$.  On the other hand, having a minimal set of primitive vectors is necessary in order to compute the volume $\Delta V^*$ of the unit cell [see, e.g., Eqs.~(\ref{pmm}) and~(\ref{F^Pi})].  
To this purpose, we use the Smith normal form of $\bf{K}$~\cite{marcus}: 
\begin{equation}\label{eq:smith}
{\bf K}={\bf T}{\bf A}{\bf P}{\bf S},
\end{equation}
where ${\bf T} \in \mathbb{Z}^{r\times r}$ and ${\bf S} \in{\mathbb Z}^{(d-1)\times (d-1)}$ are unimodular matrices, i.e. invertible matrices over the integers with~$\det{\bf T}=\det{\bf S}=\pm 1$; ${\bf A}\in \mathbb{Z}^{r\times r}$ is a diagonal matrix~with entries
\begin{equation}
\label{diag A}
({\bf A})_{ll}={\gcd \{[{\bf K}]_l\} \over \gcd \{[{\bf K}]_{l-1}\} },
\end{equation}
where $\gcd$ stands for greatest common divisor and $\{[{\bf K}]_l\}$ is the set of all minors of $\bf K$ of order~$l$;
$\bf P$ is the $r\times (d-1)$ matrix with entries
\begin{equation}\label{eq:projectionmap}
({\bf P})_{ll}=1,\quad  1\le l\le  r,
\end{equation}
and zero otherwise. 
Using~(\ref{eq:smith}) we can define the lattice~$\widetilde{\mathscr P}_n$ in terms of the new vectors/variables 
\begin{equation}\label{smith variables}
{\bf s}=(s_1,\dots,s_r)^t:={\bf P}{\bf S}{\bf n},
\end{equation}
which we coin  \emph{Smith vectors/variables}.  Given  a matrix~${\bf K}$  the Smith form guarantees that the choice of Smiths variables is unique, up to linear combinations within the degenerate subspaces of ${\bf A}$ (if any).   
Note that since the matrix $\bf S$ (and analogously ${\bf T}$) 
is unimodular, the $\gcd$ of each of its rows and columns is unity. Indeed, for the first column (similarly for the other columns/rows) one has~${\bf S}_1=\gcd\{({\bf S})_{j,1}\}\,{\bf a}_1$, for~some ${\bf a}_1\in{\mathbb Z}^{d-1}$. Then, $1=\det{\bf S}=\gcd\{({\bf S})_{j,1}\}\det \bar {\bf S}$, where~$\bar {\bf S}$ is obtained by substituting ${\bf a}_1$ for the first column of~${\bf S}$. Since $\det \bar {\bf S}\in{\mathbb Z}$, necessarily $\det \bar {\bf S}=\gcd\{({\bf S})_{j,1}\}=1$.
%
%
%
%
%
%
Being the case that the $\gcd$ of each row of ${\bf S}$ is unity, Bezout lemma ensures that each component of ${\bf s}$, $s_l$,  will take {\em all} integer values. That is, in stark contrast to the variables~$\widetilde{n}_i$, the Smith variables $s_l$ take values independently of each other, with unit spacing between consecutive values. 
This means that the Bravais lattice $\widetilde{\mathscr P}_\infty$ is defined in terms of the Smith variables as
\begin{equation}
\widetilde{\mathscr P}_\infty=\mbox{$
\left\{\vecs{n}=\sum_{l=1}^{r\phantom{d}} \,s_l\,{\bf v}_l,\; s_l\in{\mathbb Z}\right\}
$},
\end{equation}
where $\{{\bf v}_l\}_{l=1}^r$ are a set of (linear independent) primitive vectors of the lattice defined by each of the $r$ columns of~${\bf T}{\bf A}$. It follows that the volume of the unit cell can be computed as 
\begin{equation}
\Delta V^*\!=\vert\det({\bf v}_1,{\bf v}_2,\dots,{\bf v}_r)\vert
=\det {\bf A}=\gcd \{[{\bf K}]_r\},
\end{equation}
where we have used Eq.~(\ref{diag A}).

\subsubsection{Parametrizing the energy in terms of the Smith variables}
Both the vectors $\widetilde {\bf n}$ and the Smith vectors~${\bf s}$ are in one-to-one correspondence with the distinct energies in~${\sf Sp}_n$, and thus they are interchangeable in all  our arguments.
As a matter of fact,   they would coincide had we chosen the energy units as 
\begin{equation}
{\vec \varepsilon}^{\rm S}:={\vec \varepsilon}\,{\bf T}\,{\bf A}
\label{epsilon^S}
\end{equation} 
(the script S stands for Smith), so that the total energy becomes
$
E_{\bf s} =\boldsymbol\varepsilon^{\rm S}\,{\bf s}
$.    Note that, if ${\vec \varepsilon}$ is minimal, then also ${\vec \varepsilon}^{S}$ must be  minimal, since the two vectors are related by an invertible matrix with integer entries.   

Despite the one-to-one correspondence,  the Smith variables  ${\bf s}$ are more convenient than the variables $\widetilde {\bf n} $.  
Indeed,  they define a cubic lattice with unit spacing, which facilitates, e.g., taking the continuum limit. 
Note that one has  
\begin{align*}
\det\widetilde {\bf\Sigma}&=\!\det({\bf K}{\bf\Sigma}{\bf K}^t)\\
  &  =\!\det\left[  ({\bf T\,A\,P\,S} )     \,  {\bf  \Sigma } \,   ({\bf T\,A\,P\,S})^t  \right] \\
  &  =\!\det\kern-.2em{}^2\!{\bf A}\!\,\det{\bf\Sigma}^{\rm S}  \qquad    {\bf\Sigma}^{\rm S}:= {\bf P\,S}\,{\bf \Sigma}({\bf P\,S})^t  \\ 
 &\equiv \!(\Delta{V^{*}}\!)^2\! \det\!{\bf\Sigma}^{\rm S}\! \, ,
\end{align*}
which implies the relation 
\begin{equation}\label{invariance}
 \Delta{V^{*}}  \,    \left( \det\widetilde {\bf\Sigma}\right)^{-\frac 12}    =        \left( \det {\bf\Sigma}^{\rm S}\right)^{-\frac 12}  \, .
\end{equation}
Using this fact,   the volume of the unit cell     in Eqs. $\eqref{pmm}$ and \eqref{F^Pi}  can absorbed into the matrix   ${\bf\Sigma}^{\rm S}$, which  is the covariance matrix of the probability distribution of Smith variables $p_{{\bf s},n}$.  

\subsubsection{Equivalence of different choices of energy units}

It has already been mentioned that the choice of rationally independent energy units that define ${\bf K}$ is not unique. Here we show that such ambiguity has no physical implications.  Our strategy is to show that different choices of energy units lead to Smith variables that are related by  unimodular matrices.    

Let  $\boldsymbol{\varepsilon}$ and $\boldsymbol{\varepsilon}'$ be two minimal sets of rationally independent energy units spanning the spectrum of $H$.  In the following we will use the same notation of the previous section,  attaching primes to all the matrices and quantities defined in terms of $\boldsymbol{\varepsilon}'$.   Note that, by minimality, there must be an invertible transformation ${\bf R}$ such that~$\vec \varepsilon= \vec \varepsilon'\,{\bf R}$. 
Comparing the two relations     $E_{\bf n}=\boldsymbol{\varepsilon}'\,{\bf K}'\,{\bf n}$ and $E_{\bf n}=\boldsymbol{\varepsilon}\,{\bf K} \, {\bf n}  =  \boldsymbol{\varepsilon}'\,{\bf R\,  K}\,{\bf n}$   and using the rational independence of the units $\vec \varepsilon'$ we obtain  the relation 
\begin{align}
\nonumber {\bf K'}\,{\bf n}  &=  {\bf R}\,{\bf K}\,{\bf n}  \\
  \nonumber &  =     {\bf R}\,{\bf T}\,{\bf A}\,{\bf P}\,{\bf S}\,{\bf n}  \\
 \nonumber &  =    {\bf R}\,{\bf T}\,{\bf A}\,{\bf s}    \\
 \label{a}   &  =          {\bf M}\,{\bf s},   \qquad     {\bf M}  :  =  {\bf R}\, {\bf T}\,{\bf A}            \, ,
\end{align}
the second and third equalities coming from the  Smith form in Eq.~(\ref{eq:smith}) and the definition of the Smith vectors in Eq.~(\ref{smith variables}), respectively.   Now, recall that  $\bf K'$ is a matrix of integers, and, therefore ${\bf K}'\, {\bf n}$  is a vector of integers for every ${\bf n}\in{\mathbb Z}^{d-1}$.  
 Since the Smith variables $s_l$ are independent and take all possible integer values, by choosing ${\bf s}=(1,0,\dots,0)^t$, the relation ${\bf K}'\, {\bf n}  =            {\bf M}\,{\bf s} $  implies that the first column of ${\bf M}$ must have integer entries. By the  same argument, all columns of ${\bf M}$ must have integer entries, i.e.   $     {\bf M}   $ is an integer matrix.   
 Note that $\bf M$ is invertible (since it is defined as the product of three invertible matrices), although its inverse needs not be a matrix with integer entries.  

Let us write    ${\bf M}$ in the Smith form ${\bf M}={\bf U}\,  {\bf B} \,{\bf V}$, where~${\bf U}$ and ${\bf V}$ are unimodular and $\bf B$ is an invertible diagonal matrix.  Inserting this expression in Eq. (\ref{a}), we obtain  
\begin{align*}  {\bf K}' \, {\bf n }   &  =   {\bf U}\, {\bf B}\, {\bf V} \, {\bf s}     \\
&  =      {\bf U}\, {\bf B}\, {\bf V} \, {\bf P} \,  {\bf S} \,  {\bf n} \, ,
\end{align*}  
having used Eq. (\ref{smith variables}). Since the relation holds for arbitrary integer vectors, we obtained    ${\bf K}'={\bf U}{\bf B}{\bf V}{\bf P}{\bf S}$.  Note that, by definition of the matrix $\bf P$  in Eq. (\ref{eq:projectionmap}), we have 
\begin{equation}\label{aa}    {\bf V \, P}  =  {\bf   P  \,  W} \, ,
\end{equation} where $\bf W$ is the $(d-1)\times (d-1)$ matrix defined as, e.g., ${\bf W}  =  {\bf V}  \oplus   {\boldsymbol\openone}_{d-1-r}$.    Hence, we have 
${\bf K}'={\bf U}{\bf B}{\bf P}{\bf W}{\bf S}$.
Now, comparing this equation with the Smith form  ${\bf K}'={\bf T}'{\bf A'}{\bf P'}{\bf S'}$ we  obtain  $  {\bf T}'  = {\bf  U}$,  ${\bf A'}  =   {\bf B}$, ${\bf P'}  ={\bf P} $ and ${\bf S}'  =  {\bf W  \,  S}$.    In conclusion, the Smith variables defined by $\bf K'$ are related to the Smith variables defined by $\bf K$ through the relation  
\begin{align*}
\bf s'     & \equiv   {\bf P'   \,  S'  \,  n}\\
   &  =       {\bf P   \,  W\,    S  \,  n} \\
   &  =     {\bf V   \,  P\,    S  \,  n} \\ 
   &  \equiv   {\bf V   \, s}  \, ,
\end{align*}
having used Eq. (\ref{aa}) in the third equality. 
Clearly,    the covariance matrix of $\bf s'$ is given by ${\bf \Sigma}^{\rm S'}  =  {\bf V}  {\bf \Sigma}^{\rm S}   {\bf V}^t $, where~$  {\bf \Sigma}^{\rm S}$  is the covariance matrix of $\bf s$, and, thanks  to the unimodularity of $\bf V$,  we  have $\det  {\bf \Sigma}^{\rm S'}    =  \det  {\bf \Sigma}^{\rm S} $.    Using Eq. (\ref{invariance}) we  conclude that  
 \begin{align*}
 \Delta{V}^{*}(\det\widetilde{\bf \Sigma})^{-1/2}  &=(\det{\bf \Sigma}^{\rm S})^{-1/2} \\
   &  =(\det{\bf \Sigma}^{\rm S'})^{-1/2}  \\
   &    =   \Delta{V}^{* \prime}(\det\widetilde{\bf \Sigma}')^{-1/2}        \, .
 \end{align*} 
  This proves that physically relevant quantities, such as the most probable energy in Eq.~\eqref{pmm} or  the fidelity \eqref{F^Pi},  do not depend on the choice of energy units used to represent the spectrum of $H$.

\subsubsection{Boundary defects}

 Ideally, one would like to have $\widetilde{\mathscr P}_n=\widetilde{\mathscr P}_\infty\cap\widetilde\Delta^r_n$, however, near the boundary of the polytope $\widetilde\Delta^r_n$ some sides are missing, giving rise to loss of regularity, 
 as already mentioned. We argue here that the thickness of the defective region does not scale with $n$. This ensures that in the asymptotic limit of large $n$, e.g., the sums over $\widetilde{\mathscr P}_\infty$~can always be safely approximated by integrals over~$\widetilde\Delta^r_n$, even when the probability distribution $p_{{\bf n},n}$ is flat, as required to obtain Eq.~(\ref{low P fide}). 
The thickness of the defective region depends solely on the intrinsic properties of the Hamiltonian $H$, 
including the number of rationally independent energy units, through the matrix~${\bf K}$ . 

To simplify the argument let us assume that $\widetilde{\mathscr P}_\infty$ is a cubic lattice with unit spacing. It has been shown above that this assumption does not entail any loss of generality, as it just requires choosing energy units as in~(\ref{epsilon^S}). Let~${\bf K}^l=(k_{l\,1},\dots,k_{l\,d-1})$ stand for the row $l$ of the matrix~${\bf K}$. Then,
for each $l=1,\dots,r$, 
\begin{equation}
\widetilde n_l={\bf K}^l\cdot{\bf x},\quad {\bf x}\in{\mathbb R}^{d-1}, 
\label{defect1}
\end{equation}
where $\widetilde n_l$ is the $l$-th component of $\widetilde{\bf n}\in\widetilde{\mathscr P}_\infty\cap\widetilde\Delta^r_n$, defines a discrete family of hyperplanes  that are orthogonal to~${\bf K}^l$. The kernel of ${\bf K}\in{\mathbb Z}^{r\times(d-1)}$ can always be spanned  by a set $\{{\bf u}_k\}_{k=1}^{d-r-1}$ of independent vectors of ${\mathbb Z}^{d-1}$ that are orthogonal to $\{{\bf K}^l\}_{l=1}^r$. Upper bounds, $b$, to the minimum length of these (integer) vectors  (high-dimensional extensions of the so-called Siegel's bound) can be found in~\cite{bombieri,borosh} and depends entirely on the matrix ${\bf K}$ ($b$ is thus independent of $n$). 
We note that $\{{\bf u}_k\}_{k=1}^{d-r-1}$ define a lattice within $\ker{\bf K}$ and any $(d-r-1)$-dimensional ball or radius $b$ (or larger) contains at least one site of it. 

A site $\widetilde{\bf n}\in\widetilde{\mathscr P}_\infty\cap\widetilde\Delta^r_n$ belongs to $\widetilde{\mathscr P}_n$ iff there is at least one site ${\bf n}\in{\mathscr P}_n$ that satisfies~(\ref{defect1}) for all $l$. The Smith normal form and Bezout lemma (see previous subsection) ensure that there are infinitely many vectors~${\bf n}$ in ${\mathscr P}_\infty$ satisfying~(\ref{defect1}) for a given $\widetilde{\bf n}$ (but they are not necessarily in~$\widetilde{\mathscr P}_n$). The difference between any two such vectors, ${\bf n}-{\bf n}'$ belongs to $\ker{\bf K}$, i.e., satisfies ${\bf n}-{\bf n}'=\sum_{k=1}^{d-r-1}\eta_k{\bf u}_k$, $\eta_k\in{\mathbb Z}$. Therefore, if a given $\widetilde{\bf n}$ is such that the intersection of the polytope $\widetilde\Delta^r_n$ with the hyperplanes defined by~(\ref{defect1}) contains a ball of radius $b$, it is ensured that $\widetilde{\bf n}\in\widetilde{\mathscr P}_n$. This is so because at least one site in ${\bf n}_0+\ker{\bf K}$, where~${\bf n}_0\in {\mathscr P}_\infty$ is a solution (any of the infinitely many solutions) of~(\ref{defect1}), will be contained in this ball, as argued above. Since~$\Delta^{d-1}_{c,n}$ is convex, by increasing $n$ (the length of its edge) the volume of its intersection with the hyperplanes~(\ref{defect1}), which is also a convex polytope, grows as $n^{d-r-1}$ and it will eventually contain a ball of radius~$b$. Only sides whose distance to the boundary is kept fixed and is small enough can escape from this fate and may thus not be in $\widetilde{\mathscr P}_n$. This concludes our proof.

By the same argument, since the volume of the intersection of $\Delta^{d-1}_{c,m}$ with the hyperplanes~(\ref{defect1}) grows with~$m$ (but at a fixed distance from the boundary), the number of  balls or radius $b$ it will eventually contain grows as~$m^{d-r-1}$ and so does the number of point in $\ker{\bf K}$, i.e., the numbers of points in ${\mathscr P}_m$ associated to a given $\widetilde{\bf m}$, as required to obtain, e.g.,  Eq.~(\ref{pmm}).

As a final remark, we note also that the pattern of defects near the boundary of $\tilde\Delta^r_n$ is exactly the same for every $n$, provided that $n$ is sufficiently large. The reason  is that all the lattices ${\mathscr P}_n$ are similar (i.e., have the same shape but different size). Hence, the regions of them that are at a fixed distance from the boundary are identical and, as argued above, these regions determine the pattern of defects of $\widetilde{\mathscr P}_n$.

\subsection{Equivalence between PM and macroscopic cloning of quantum clocks}

In this section we show that the PM fidelity attains the upper bound~(\ref{upper bound}) to the cloning fidelity, and thus prove the equivalence between PM and macroscopic cloning of quantum clocks. The proof uses the one-to-one correspondence between ${\sf Sp}_m$ and $\widetilde{\mathscr P}_m$, as well as the fact that the probability distribution $p_{\widetilde{\bf m},m}$ approaches a multivariate Gaussian distribution as $m$ goes to infinity. The former, enables us to write Eq.~(\ref{F_M&P}) as
\begin{equation}
F^\Pi_{\rm PM}=\sum_{\widetilde\mus}
\left(\sum_{\widetilde{\bf n}\in\widetilde{\mathscr P}_n}\kern-.5em\raisebox{.2em}{$'$}\xi_{\widetilde{\bf n}}\sqrt{\hat p_{\widetilde{\bf n}+\widetilde\mus,m} \, p_{\widetilde{\bf n}+\widetilde\mus,m}}\right)^2,
\end{equation}
where $\xi_{\widetilde{\bf n}}:=\sqrt{p_{\widetilde{\bf n},n}\pi_{\widetilde{\bf n}}/P_{\rm succ}}$ and the outer sum runs over $\widetilde{\mathscr P}_{n,m}:=\{\widetilde{\boldsymbol{\mu}}=\widetilde{\bf m}-\widetilde{\bf n}\;|\; \widetilde{\bf n}\in\widetilde{\mathscr P}_n, \widetilde{\bf m}\in\widetilde{\mathscr P}_m\}$. To keep our notation as uncluttered as possible we suppress the tildes throughout the rest of this section.
We further assume that the multivariate normal distribution $p_{{\bf m},m}$ peaks at~${\bf m}={\bf 0}$, and so does $p_{{\bf n},n}$, but we make no other assumption on the form of $p_{{\bf n},n}$ (it could, e.g., be flat, in which case any point in ${\mathscr P}_n$ could be chosen to be~${\bf 0}$). This may require shifting the vectors in~${\mathscr P}_m$ by a fixed~${\bf m}_0\in{\mathscr P}_m$ (similarly, by a fixed~${\bf n_0}\in{\mathscr P}_n$ for those in~${\mathscr P}_n$).
The primed summation is restricted to vectors~$\bf n$ such that {\boldmath$\mu$}+{\bf n}$\in {\mathscr P}_m$.

For any $\rho>\nu>0$, where $\nu=\max\{|{\bf n}|: {\bf n}\in{\mathscr P}_n\}$, define the set 
${\cal R}_\rho=\{\mbox{\boldmath$\mu$}: \forall{\bf n}\!\in\!{\mathscr P}_n,\,  \mbox{\boldmath$\mu$}+{\bf n}\!\in\!{\mathscr P}_m\, \mbox{and}\ |\mbox{\boldmath$\mu$}+{\bf n}|\!\le\!\rho\}$. Then
\begin{equation}
F^\Pi_{\rm PM}>\sum_{\mbox{\boldmath$\scriptstyle \mu$}\in{\cal R}_\rho}
\left(\sum_{\bf n}\xi_{\bf n}\sqrt{\hat p_{{\bf n}+\mus,m} p_{{\bf n}+\mbox{\boldmath$\scriptstyle \mu$},m}}\right)^2 .
\end{equation}
Note that we can drop the prime in the last sum over~${\bf n}$. We have
\begin{equation}
F^\Pi_{\rm PM}>
p_{{\bf 0},m}\; {\rm e}^{-{ \rho^2\over 2m\sigma_1^2}} \sum_{\mbox{\boldmath$\scriptstyle \mu$}\in{\cal R}_\rho}
\left(\sum_{\bf n}\xi_{\bf n}\sqrt{\hat p_{{\bf n}+\mus,m}} \right)^2,
\end{equation}
where $m \sigma_1^2$ is the smallest eigenvalue of the covariance matrix of $p_{{\bf m},m}$. Here, we explicitly display the $m$ dependence of the covariance matrix eigenvalues; thus $\sigma_1^2$ does not scale with $m$. Let us choose the `guessed' distribution~as
\begin{equation}
\hat p_{{\bf m},m}=\hat p_{{\bf 0},m}\;{\rm e}^{-\zeta {|{\bf m}|^2\over 2m}};\quad
\sum_{{\bf m}\in{\mathscr P}_m} \hat p_{{\bf m},m}=1.
\end{equation}
%
Then,
\begin{eqnarray}
&&\kern-1.5em \hat p_{{\bf n}+\mbox{\boldmath$\scriptstyle\mu$},m}=\hat p_{{\bf 0},m}\;{\rm e}^{-\zeta {|{\bf n}+\mbox{\boldmath$\scriptscriptstyle\mu$}|^2\over 2m}}\ge
\hat p_{{\bf 0},m}\;{\rm e}^{-\zeta {(|{\bf n}|+|\mbox{\boldmath$\scriptscriptstyle\mu$}|)^2\over 2m}}\nonumber\\
&&\kern-1.5em\phantom{\hat p_{{\bf n}+\mbox{\boldmath$\scriptstyle\mu$},m}}=
{\rm e}^{-\zeta {|{\bf n}|^2+2|{\bf n}||\mbox{\boldmath$\scriptscriptstyle\mu$}|\over 2m}}
\hat p_{\mbox{\boldmath$\scriptstyle\mu$},m}
\ge
{\rm e}^{-\zeta {3|{\bf n}|^2+2|{\bf n}|\rho\over 2m}}
\hat p_{\mbox{\boldmath$\scriptstyle\mu$},m}  ,
\end{eqnarray}
where we have used that $|\mbox{\boldmath$\mu$}|\le|{\bf n}|+\rho$ if $\mbox{\boldmath$\mu$}\in{\cal R}_\rho$. Thus, the following bound holds
%
%
%
\begin{equation}
F^\Pi_{\rm PM}>
p_{{\bf 0},m}\,{\rm e}^{-{ \rho^2\over 2m\sigma_1^2}-\zeta {3\nu^2+2\nu \rho\over 2m}} 
\left(\!\sum_{\bf n}\xi_{\bf n} \!\right)^2\!\!\sum_{\mbox{\boldmath$\scriptstyle \mu$}\in{\cal R}_\rho}\!\hat p_{\mus,m}.
\end{equation}
%
We now need to lower bound the last sum.  For this, we write
\begin{equation}
\sum_{\mbox{\boldmath$\scriptstyle \mu$}\in{\cal R}_\rho} \hat p_{\mus,m}=1-\kern-.9em\sum_{\mbox{\boldmath$\scriptstyle \mu$}\in\overline{\cal R}_\rho\cap{\mathscr P}_m}\kern-.5em \hat p_{\mus,m} .
\end{equation}
For $\mbox{\boldmath$\mu$}\in\overline{\cal R}_\rho\cap{\mathscr P}_m$ one has $\rho<|{\bf n}+\mbox{\boldmath$\mu$}|\le\nu+|\mbox{\boldmath$\mu$}|$, then, recalling that $\rho> \nu$,
\begin{eqnarray}
\sum_{\mbox{\boldmath$\scriptstyle \mu$}\in{\cal R}_\rho} \hat p_{\mbox{\boldmath $\scriptstyle \mu$},m}&>&1-\kern-.9em\sum_{\mbox{\boldmath$\scriptstyle \mu$}\in\overline{\cal R}_\rho\cap{\mathscr P}_m}\kern-.5em \hat p_{{\bf 0},m}\;{\rm e}^{-\zeta {(\rho-\nu)^2\over2m}}\nonumber\\
&>& 1-|{\mathscr P}_m| \,\hat p_{{\bf 0},m}\;{\rm e}^{-\zeta {(\rho-\nu)^2\over2m}} .
\end{eqnarray}
Note that $|{\mathscr P}_m|\le (m+d-1)!/[m!(d-1)!]\sim  m^{d-1}$, for large $m$.
With all the above,
\begin{eqnarray}
F^\Pi_{\rm PM}&>&\left( \max_{\bf m} p_{{\bf m},m}\right) {\rm e}^{-{ \rho^2\over 2m\sigma_1^2}-\zeta {3\nu^2+2\nu \rho\over 2m}} 
\nonumber\\
&\times&
\left[1-C m^{d-1}{\rm e}^{-\zeta {(\rho-\nu)^2\over2m}}\right]
\left(\sum_{\bf n}\xi_{\bf n} \right)^2
\end{eqnarray}
for some positive constant $C$.
Therefore, if $\rho=m^{{1-\epsilon\over2}}$, and $\zeta=m^{\delta}$, with $(1+\epsilon)/2>\delta>\epsilon>0$, then
$\rho^2/m=m^{-\epsilon}$, $\zeta/m=m^{-1+\delta}$, $\zeta \rho/m=m^{-{1+\epsilon\over2}+\delta}$ and $\zeta \rho^2/m=m^{\delta-\epsilon}$. Thus, for large $m$ we have
\begin{equation}
F^\Pi_{\rm PM}>\left( \max_{\bf m} p_{{\bf m},m}\right)
\left(\sum_{\bf n}\xi_{\bf n} \right)^2.
\label{final lower bound}
\end{equation}
This result holds provided $p_{{\bf m},m}$ is a multivariate normal distribution picked at some ${\bf m}_0\in{\mathscr P}_m$. The actual distribution is multinomial on the points of ${\mathscr P}_m$. However, as $m$ becomes asymptotically large, the induced distribution $p_{{\bf m},m}$ becomes arbitrarily closed to the multivariate normal assumed in the proof above. Recalling that the right hand side of~(\ref{final lower bound}) is also an upper bound to $F^\Pi_{\rm CL}$ and, thus to $F^\Pi_{\rm PM}$, we finally conclude that (we restore the suppressed tildes)
\begin{equation}
F^\Pi_{\rm PM}=F^\Pi_{\rm CL}=\left( \max_{\widetilde{\bf m}\in\widetilde{\mathscr P}_m} p_{\widetilde{\bf m},m}\right)
\left(\sum_{\widetilde{\bf n}\in\widetilde{\mathscr P}_n}\xi_{\widetilde{\bf n}} \right)^2
\label{final lower bound =}
\end{equation}
for asymptotically large $m$ and fixed $n$. This leads to~Eq.~(\ref{F^Pi}), of which Eq.~(\ref{F^Pi_PM = F^Pi_CL}) is a particular case for $r=1$.

\subsection{Explicit calculations}

In this section we give some details of the calculation leading to Eqs.~(\ref{low P fide}) and~(\ref{high P fide}). As already mentioned, Smith vectors ${\bf s}\in{\mathbb Z}^r$ [recall Eq.~(\ref{smith variables})] are most suited to this purpose because they form a cubic lattice of unit step size, i.e., their minimal cell has volume $\Delta V^*=1$. In this sense, they are just a particular instance of  vectors $\widetilde{\bf m}\in\widetilde{\mathscr P}_m$. Hence, to avoid further proliferation of notation, we will use here the generic symbols  $\widetilde{\bf \Sigma}$ and~$\widetilde{\mathscr P}_m$ to refer to the covariance matrix of the multivariate Gaussian distribution $p_{{\bf s},m}$ and the lattice of the Smith vectors ${\bf s}$ respectively. Since $\widetilde{\bf\Sigma}$ scales with $m$, we write $\widetilde{\bf \Sigma}=m\widetilde{\bf \Sigma}_1$, where $\widetilde{\bf \Sigma}_1$ is independent of $m$. Then, the expression for the asymptotic fidelity in~(\ref{F^Pi}) becomes 
\begin{equation}\label{eq:ap-fide}
F^{\Pi}=
{1\over \sqrt{(2\pi m)^r\det \widetilde{\bf \Sigma}_1}}
\left(\sum_{{\bf s}\in \widetilde{\mathscr{P}}_n} \sqrt{p_{{\bf s},n}{\pi_{{\bf s}}\over P_{\rm succ}}}\right)^2 .
\end{equation}
The last sum can be evaluated in the asymptotic limit of large $n$ (recall however that we assume $n\ll m$), as we show below. 
In this case, also $p_{{\bf s},n}$ approaches a multivariate normal distribution, as  that in~Eq.~(\ref{pmm}), and the sum over $\widetilde{\mathscr P}_n$ can be approximated by an integral over the polytope~$\widetilde\Delta^r_n$.

As a warmup act, we first compute the fidelity $F^{\Pi}$ for the deterministic protocol, i.e., when $P_{\rm succ}=1$. Since no filtering is applied,  we have $\pi_{{\bf s}}=1$ for all ${\bf s}\in\widetilde{\mathscr P}_n$. The sum in~(\ref{eq:ap-fide}) simplifies to
\begin{equation}
\int_{\widetilde\Delta^r_n}\!\!\!\!d{\bf s}\,  \sqrt{  p_{{\bf s},n} }
\simeq\!\!
\int_{{\mathbb R}^r}\!\!\!\! d{\bf s}\,  \sqrt{  p_{{\bf s},n} }
=
2^{r/2}\!\left[(2\pi n)^{r} \det\widetilde{\bf\Sigma}_1\right]^{1/4}\!\!, 
\end{equation}
as $2^{-r/2}(2\pi)^{-r/4}(\det\widetilde{\bf\Sigma})^{-1/4}\times\sqrt{  p_{{\bf s},n} }$ is also a properly normalized multivariate normal distribution with covariance matrix $2\widetilde{\bf\Sigma}$. Substituting in Eq.~(\ref{eq:ap-fide}) we obtain
\begin{equation}
F=\left(4{n\over m}\right)^{r/2}.
\end{equation}
This expression agrees with Eq.~(\ref{high P fide}) in the deterministic limit, when $\eta\to0$.

In the probabilistic case, $P_{\rm succ}<1$, we need to optimize the filter parameters $\{\pi_{\bf s}\}$. To simplify the notation, let us use the definition of the normalized state $|\xi\rangle$, with components $\xi_{{\bf s}}:= \sqrt{p_{{\bf s},n}\,{\pi_{\bf s}/ P_{\rm succ}}}$. Then, the maximum fidelity, $F=\max_{\Pi}F^\Pi$, 
is obtained when the sum in~(\ref{eq:ap-fide}) takes its maximum value:
%
\begin{eqnarray}\label{eq:kkt-target}
\max & & \sum_{{\bf s}} \xi_{{\bf s}} \; ,\\
\label{eq:kkt-primal1}
\textrm{subject to }&\quad &\sum_{{\bf s}} \xi_{{\bf s}}^2=1\; \\
\label{eq:kkt-primal2}
{\rm and}& &\xi_{{\bf s}} \leq \sqrt{p_{{\bf s},n}\over{P_{\rm succ}}} ,\quad {\bf s}\in\widetilde{\mathscr P}_n,
\end{eqnarray}
where (\ref{eq:kkt-primal1}) is the normalization constraint, and (\ref{eq:kkt-primal2}) comes from the positivity and trace preserving requirements on the stochastic filter. To solve (\ref{eq:kkt-target}), (\ref{eq:kkt-primal1}) and~(\ref{eq:kkt-primal2}), we use Lagrange multipliers and the Karush-Kuhn-Tucker conditions. The problem reduces to solving the stationary conditions
\begin{eqnarray}
\nonumber 
{\partial\over\partial\xi_{{\bf s}}}\!\left(\sum_{{\bf s}'} \xi_{{\bf s}'}\!\!\right)&=&\,{\partial\over\partial\xi_{{\bf s}}}\Bigg[ {1\over2\zeta} \left(\sum_{{\bf s}'} \xi_{{\bf s}'}^2-1\right)\\
&+&\sum_{{\bf s}'} \sigma_{{\bf s}'} \left(\xi_{{\bf s}'} - \sqrt{p_{{\bf s}',n}\over{P_{\rm succ}}}\,\right)\Bigg],
\label{stat con}
\end{eqnarray}
where the sums extend to ${\bf s}\in\widetilde{\mathscr P}_n$, and $1/(2\zeta)$ and $\{\sigma_{{\bf s}}\}_{{\bf s}\in\widetilde{\mathscr P}_n}$ are multipliers. Conditions~\eqref{eq:kkt-primal1} and~\eqref{eq:kkt-primal2} are called primal feasibility conditions. In addition, one has to impose that~$\sigma_{{\bf s}}\geq 0$, known as dual feasibility condition, and
\begin{equation}
\sigma_{{\bf s}}\left(\xi_{{\bf s}} - \sqrt{p_{{\bf s},n}\over{P_{\rm succ}}}\right)=0,
\end{equation}
known as complementary slackness condition, both  for all $ {\bf s}\in\widetilde{\mathscr P}_n$.
The latter, implies that at any site of $\widetilde{\mathscr P}_n$, either $\sigma_{\bf s}=0$ or $\xi_{\bf s}=\sqrt{p_{{\bf s},n}/P_{\rm succ}}$, in which case we say that ${\bf s}$ belongs to the coincidence set $\mathscr C$, i.e., ${\mathscr C}:=\{{\bf s}\in\widetilde{\mathscr P}_n : \xi_{{\bf s}}^2=p_{{\bf s},n}/P_{\rm succ}\}$. If ${\bf s}\not\in{\mathscr C}$, Eq.~(\ref{stat con}) readily gives the constant solution $\xi_{\bf s}=\zeta$. 

If $P_{\rm succ}< \min_{{\bf s}\in\widetilde{\mathscr P}_n} p_{{\bf s},n}$, then $p_{{\bf s},n}/P_{\rm succ}> 1\ge \xi^2_{{\bf s}}$, thus~${\mathscr C}=\emptyset$. In this case, normalization implies $\zeta=|\widetilde{\mathscr P}_n|^{-1/2}$, where $|\widetilde{\mathscr P}_n|$ is the number of sites in $\widetilde{\mathscr P}_n$. Substituting in~(\ref{eq:ap-fide}), we have
\begin{equation}
F={\vert\widetilde{\mathscr P}_n\vert\over \sqrt{(2\pi m)^r\det\widetilde{\bf\Sigma}_1}}\;.
\label{low P fide2}
\end{equation}
Recalling Eq.~(\ref{invariance}) and $m\widetilde{\bf\Sigma}_1=\widetilde{\bf\Sigma}$, this equation becomes Eq.~(\ref{low P fide}), which holds for any choice of vectors $\widetilde{\bf n}$. Notice that both, Eqs.~(\ref{low P fide2}) and~(\ref{low P fide}), also hold for small~$n$. For large $n$, $\vert\widetilde{\mathscr P}_n\vert \Delta V^*$  (recall $\Delta V^*=1$ for Smith vectors) approaches $V_n$, the volume of the polytope $\tilde\Delta^r_n$, as the irregularities or defects of the lattice $\widetilde{\mathscr P}_n$ can only arise within a finite distance from its boundary (see previous sections). Closed formulas for  $\vert\widetilde{\mathscr P}_n\vert $ or $V_n$ depend on the Hamiltonian $H$ and do not seem to generalize easily. In the main text, only the obvious scalings $\vert\widetilde{\mathscr P}_n\vert\sim V_n\sim n^r$ are used to show that $F\sim (n/\sqrt m)^r$.

\begin{figure}[htbp]
\begin{center}
\includegraphics[width=17em]{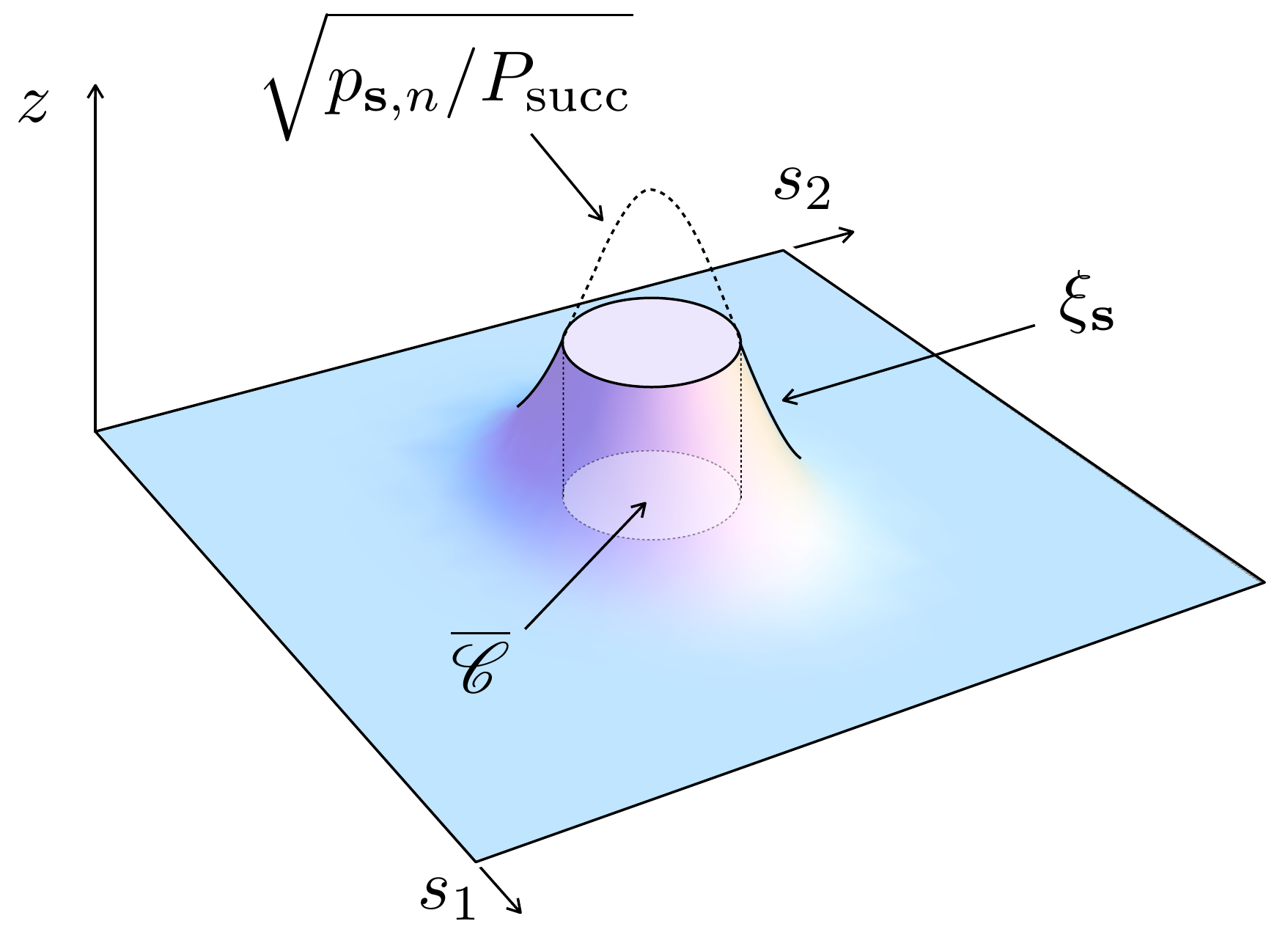}
\caption{\label{fig1}Plot of $\xi_{{\bf s},n}$  ($z$ axis) for large $n$ (the truncated bell-shaped surface). The figure also shows the complement of the coincidence set and~$\sqrt{p_{{\bf s},n}/P_{\rm succ}}$ for~$P_{\rm succ}>\min_{{\bf s}\in\widetilde{\mathscr P}_n} p_{{\bf s},n}$.}
\end{center}
\end{figure}
For $P_{\rm succ}> \min_{{\bf s}\in\widetilde{\mathscr P}_n} p_{{\bf s},n}$ the problem becomes more involved, as the constraint~(\ref{eq:kkt-primal2}) is now non-trivial and~${\mathscr C}\not=\emptyset$. Since $p_{{\bf s},n}/P_{\rm succ}$ is bell-shaped, the complement of the coincidence set is the `ellipsoid' $\overline{\mathscr C}_\alpha:=\{{\bf s}\in\widetilde{\mathscr P}_n : ({\bf s}-{\bf s}_0)^t\widetilde{\bf\Sigma}^{-1}({\bf s}-{\bf s}_0)\le\alpha^2$, for some $\alpha$ (Fig.~\ref{fig1} shows a two-dimensional version of it), and we have the solution 
 \begin{equation}
\xi_{{\bf s}}=
\begin{cases}
\sqrt{p_{\alpha,n}/{P_{\rm succ}}}&\text{if}\quad {\bf s}\in\overline{\mathscr C}_\alpha
\\[.7em]
\sqrt{p_{{\bf s},n}/{P_{\rm succ}}} &\text{if}\quad  {\bf s}\not\in\overline{\mathscr C}_\alpha\;,
\end{cases}
\label{sol xi}
\end{equation}
where we have defined
\begin{equation}
p_{\alpha,n}:={\expo{-\alpha^2 / 2}\over \sqrt{(2\pi)^r\det \widetilde{\bf\Sigma}}},
\end{equation}
and we note that the parameter $\alpha$ that gives the size of the `ellipsoid' $\overline{\mathscr C}_\alpha$ is determined by normalization: $1=\sum_{\bf s}\xi_{\bf s}^2\simeq \int d^r\!{\bf s}\,\xi_{\bf s}^2$. 
Note also that the solution $\xi_{\bf s}$ is a `truncated' multivariate normal distribution, as shown in Fig.~\ref{fig1}. The above integral thus splits into two straightforward ones over the two regions in~(\ref{sol xi}), and we obtain the relation:
\begin{equation}
P_{\rm succ}={1\over \Gamma({r\over 2})}
\left(\Gamma(\mbox{\normalsize{${r\over 2}$,${\alpha^2\over 2}$}})+{\alpha^r \expo{-\alpha^2/2}\over 2^{r/2-1}r}\right),
\label{psucc}
\end{equation}
where $\Gamma(a,x)$ is the upper incomplete Gamma function. Proceeding along the same line, one can compute $\sum_{\bf s}\xi_{\bf s}$ to obtain
\begin{equation}
F\!=\!{1\over \Gamma({r\over 2}\!+\!1)}\!\left({n\over 2 m}\right)^{\!r/2}\!\!{\left[2^{r-1}r\Gamma({r\over2},\!{\alpha^2\over 4})\!+\!\alpha^r\expo{-\alpha^2\!/4}\!\right]^{\!2}\over 2^{r/2-1} r\Gamma({r\over 2},\!{\alpha^2\over 2})\!+\!\alpha^r\expo{-\alpha^2\!/2}}.
\label{explicit fide} 
\end{equation}
%
Eqs.~(\ref{psucc}) and~(\ref{explicit fide}) give the solution to our optimization problem in parametric form, in terms of $\alpha$.
Finding the fidelity $F$ as a explicit function of $P_{\rm succ}$ would require inverting the relation~(\ref{psucc}), which cannot be done analytically for an arbitrary success probability. We can however obtain an analytic expression of $F$ for a success probability close to one, i.e., for 
$P_{\rm succ}=1-\eta$, $\eta\ll1$, by simply expanding the right hand side of Eq.~(\ref{psucc}) to leading order in $\alpha$. Such expansion reads:
\begin{equation}
\eta= 1-P_{\rm succ}={ \alpha^{2 + r}\over2^{1+r/2}\;\Gamma(2 + r/2)}+{\cal O}(\alpha^{r+3}).
\end{equation}
Solving for $\alpha$ and substituting in the expansion (at leading order in $\alpha$) of the right hand side of Eq.~(\ref{explicit fide}) one obtains
\begin{equation}
F=
\left(4{n\over m}\right)^{r/2} \left[1+(1-2^{-r/2})\;\eta\right]+{\cal O}(\eta^{2}).
\label{fide eta}
\end{equation}
Eqs.~(\ref{psucc}) through~(\ref{fide eta}) hold provided $P_{\rm succ}$ does not exponentially vanish with $n$. If it does, as in~(\ref{low P fide}), where we assumed that $P_{\rm succ}< \min_{{\bf s}\in\widetilde{\mathscr P}_n} p_{{\bf s},n}$, we obtain the better scaling $F\sim (n/\sqrt{m})^r$. The reason for this different scaling is that a multivariate normal distribution only approximates $p_{{\bf s},n}$ accurately around its peak, whereas it falls off exponentially with $n$ at the tails (where~$ \min_{{\bf s}\in\widetilde{\mathscr P}_n} p_{{\bf s},n}$ lies).

\section{Proof of theorem 1}
  
{\bf Proof.}  The proof follows the same lines of the proofs of theorems 4 and 5 in Ref. \cite{tqc2010S}. 
Let $\map C_m$ be a quantum operation transforming states on $\mathcal H_{\rm in}$ into sates on $\mathcal H^{\otimes m}$.   
 First, we suppose that the output states of $\map C_m$ have support contained in the symmetric subspace of $\mathcal H^{\otimes m}$.    
 In this case, it is useful to consider the universal measure-and-prepare channel   $ \mathcal{M}$ defined by 
 \begin{align*}
 \map{M}  (\rho) =  \int  {\rm d} \psi~d_m^+  \,  \tr [   \psi^{\otimes m} \rho]  ~   \psi^{\otimes m} 
 \end{align*}
 where ${\rm d} \psi$ is the invariant measure over the pure states,  $d_m^+  =   \binom{ m + d-1}{d-1}$ is the dimension of the symmetric subspace. Note that the map $ \map{M} $ is trace-preserving for all states with support in the symmetric subspace. Moreover, $ \map{M} $ provides a good approximation of the partial trace \cite{tqc2010S}:  
\begin{align}\label{motherapprox}
\left  \|   {\tr}_{m-k}    -    {\tr}_{m-k}   \circ \map{M} \right\|_\diamond  \le  \frac{2kd}m \, ,  
\end{align} 
where $\|   \map L \|_\diamond  $  denotes the diamond norm of a linear, Hermitian-preserving map $\map L$, defined as 
$$  \|  \map L \|_\diamond : =   \sup_{r\in\mathbb N}    \,  \sup_{ 
\begin{array}{c} 
|\Psi\rangle  \in  \mathcal H_{\rm in}  \otimes   \mathbb C^r  \\
\|  |\Psi\rangle  \|  = 1
\end{array}}  \,   {\tr}  \left |  (\map L  \otimes \map I_r)  (  |\Psi\rangle\langle  \Psi|) \right | \, ,$$
$\map I_r$ denoting the identity map for an $r$-dimensional quantum system. 
Using Eq. (\ref{motherapprox}), it is immediate to construct the desired  PM protocol.  The protocol consists in performing the quantum operation $\map C_m$ and subsequently applying the measure-and-prepare channel $\map M$. Mathematically, it is described by the  quantum operation  $\widetilde{\map C}_m  =    \map{M} \circ  \map C_m$.  Since $\map M$ is trace-preserving (in the symmetric subspace), one has 
$$   {\rm tr}   [       \widetilde{\map C}_m  (\rho) ]  =  {\rm  tr} [       {\map C}_m   (\rho)]$$
for every input state $  \rho$,  that is, $ \widetilde{\map C}_m$ and $\map C_m$ have the same success probability.   Moreover, one has the bound 
\begin{align}
\nonumber &\left\|  {\tr}_{m-k}  \circ \map C_m  -  {\tr}_{m-k}  \circ \widetilde {\map C}_m  \right\|_{\diamond}      \\  
\nonumber  &  \qquad \qquad   \le   \|  \left(   {\tr}_{m-k}    -    {\tr}_{m-k}   \circ \map{M} \right) \map C_m \|_\diamond \\
\nonumber  &  \qquad \qquad \le   \|     {\tr}_{m-k}    -    {\tr}_{m-k}   \circ \map{M}  \|_\diamond   \|  \map C_m  \|_{\diamond}\\
&  \qquad  \qquad \le  \frac{2kd}m \, , \label{defibound}
\end{align}  
having used the fact that $\|  \map C_m  \|_\diamond\le 1$ by  definition.  
In words, the $k$-copy restrictions of $\widetilde{\map C}_m$ and $\map C_m$ are close to each other provided that $k  \ll m$.  

Now, suppose that the output of  $\map C_m$ has support outside the symmetric subspace. In this case, the invariance under permutations implies that  $\map C_m$ has a Stinespring dilation of the form 
\begin{align*}
\map C_m    =    {\tr}_{\mathcal H_E}  \circ   \map K_m \qquad  \map K (\rho)  : = K_m \rho K_m^\dag,
\end{align*}
where $\mathcal H_E  =  \mathcal H^{\otimes m}$ and $K_m $ is an operator with range contained in the symmetric subspace of $\mathcal H^{\otimes m} \otimes \mathcal H_E \simeq  (  \mathcal H \otimes \mathcal H)^{\otimes m}$  (for a proof see e.g. \cite{tqc2010S}).   Hence, we can take the measure-and-prepare quantum operation $\widetilde {\map K}_m: =   \map M_E \circ \map K_m$, where $\map M_E$ is the universal measure-and-prepare channel on $\mathcal H^{\otimes m} \otimes \mathcal H_E$,   and we can define $\widetilde {\map C}_m :=  {\tr}_{\mathcal H_E}  \circ  \widetilde{\map K}_m$.  By definition, the success probability of $\widetilde {\map C}_m$ is equal to the success probability of $\map C_m$.     Moreover, one has the relation 
\begin{align*}
&\left\|  {\tr}_{m-k}\circ \map C_m  - {\tr}_{m-k}  \circ \widetilde{\map C}_m  \right\|_\diamond \\
& \qquad \qquad  =      \left\| {\tr}_{m-k}    \circ {\tr}_{\mathcal H_E}   \circ  \left  (  \map K_m  -  \widetilde{\map K}_m   \right) \right\|_\diamond    \\  
& \qquad \qquad  =    \left\|    {\tr}_{\mathcal H^k_E}    \circ  \left(  {\tr}_{m-k}  \otimes   {\tr}_{\mathcal H^{m-k}_E}  \right)  \circ  \left  (  \map K_m  -  \widetilde{\map K}_m   \right) \right\|_\diamond    
\end{align*}
where  $\tr_{\mathcal H_E^k}$ denotes the partial trace over $k$ ancillary Hilbert spaces.  
Hence, one gets  the bound 
\begin{align} 
\nonumber &\left\|  {\tr}_{m-k}\circ \map C_m  - {\tr}_{m-k}  \circ \widetilde{\map C}_m  \right\|_\diamond  \\
\nonumber & \qquad \qquad  \le    \left\| {\tr}_{\mathcal H^k_E} \right \|_\diamond   \left \|  \left(  {\tr}_{m-k}  \otimes   {\tr}_{\mathcal H^{m-k}_E}   \right)  \left (  \map K_m  -  \widetilde{\map K}_m \right)   \right\|_\diamond    \\  
\label{diamond} &\qquad \qquad \le   \frac{  2 d^2 k}m \, ,    
\end{align}
having used    Eq. (\ref{defibound}) with $\widetilde {\map C}_m$,  $\map C_m$, and $d$ replaced by  $\widetilde{\map K}_m$, $\map K_m$, and $d^2$, respectively. 

Finally, the error probability in distinguishing  between $\widetilde {\map C}_m$ and $\map C_m$ by inputting a state $\rho$ and measuring $k$ output system is equal to the error probability in distinguishing between the two states  
$$   \widetilde \rho_{m,k}  =  \frac{  {\tr}_{M-k}  [\widetilde {\map C}_m (\rho)]}{\tr[ \widetilde {\map C}_m (\rho) ]}   \quad {\rm and}\quad      \rho_{m,k}  =  \frac{  {\tr}_{M-k}  [{\map C}_m (\rho)]}{\tr[  {\map C}_m (\rho) ]}  \, ,$$
respectively. Assuming equal prior probabilities for the two quantum operations, Helstrom theorem gives the bound   $p_{\rm err}    = \frac  12  \left[    1  +   \frac 1 2  \| \widetilde \rho_{m,k}   -   \rho_{m,k}   \|_1  \right]$ and therefore we have
\begin{align*}
  p_{\rm err} &  \le    \frac  12  \left[     1  +   \frac { \|  {\tr}_{m-k}   \circ   (\widetilde{ \map C}_m  -  \map C_m  )  \|_\diamond }{2  P_{\rm succ}  (\rho)}     \right]   \\
  &   \le        \frac  12  \left[    1  +   \frac {   kd^2}{  m  P_{\rm succ}  (\rho)}     \right]  \, ,
\end{align*}
where $P_{\rm succ} (\rho)  :  =  {\tr}[  \map C_m (\rho)]  \equiv  {\tr}[ \widetilde {\map C}_m] \, .$
$\blacksquare$  

\medskip 

\section{Approximation of the optimal $k$-copy cloning fidelity}

Suppose that we are given $n$ copies of the state $|\psi_x\rangle  \in  \mathcal H$, $x\in\sf X$ and that we want to produce $m$ approximate copies, whose quality is assessed by checking a random group of $k$ output systems.  Let $\map C_{n,m}$ be the quantum operation describing the cloning process.     Since the $k$ systems are chosen at random, we can restrict our attention to quantum operations that are invariant under permutation of the output spaces.     Conditional on the occurrence of the quantum operation $\map C_{n,m}$ and on preparation of the input $|\psi_x\rangle$, the   $k$-copy cloning  fidelity is given by 
\begin{align}  
\label{kfid}
F_{k,x} [\map C_{n,m}] &=    \frac{ O_{k,x}  [\map C_{n,m}]  }{P_{x}  [\map C_{n,m}]}  \\
\nonumber O_{k,x}  [\map C_{n,m}]  & : =      \bra{\varphi_x}^{\otimes k}    {\tr}_{m-k}  [ \map C_{n,m} \left (\varphi_x^{\otimes n}      \right)]          \ket{\varphi_x}^{\otimes k}\\
\nonumber P_{x}  [\map C_{n,m}]  & : =     {\tr}  [ \map C_{n,m} \left (\varphi_x^{\otimes n}      \right)]    \, ,       
\end{align} 
where $ \psi_x$ denotes the rank-one projector $\psi_x :  =  |\psi_x  \rangle\langle \psi_x|$. 
Now,  constructing the  quantum operation $\widetilde{\map C}_{n,m}$ as in theorem 1 and using Eq. (\ref{diamond}) we  have  
\begin{align*}
\left|  O_{k,x} [ \map C_{n,m} ]    -  O [\widetilde{ \map C}_{n,m} ]   \right |  &\le  \frac{ 2k d^2} m\\
    P_{x} [ \map C_{n,m}]   &  =      P_{x} [\widetilde{ \map C}_{n,m} ]   \equiv P_{\rm succ}  \left(\psi_x^{\otimes n}\right) ,
\end{align*}
and, therefore, 
\begin{align*}  \left|     F_{k,x} [\map C_{n,m}]  -  F_{k,x} [\widetilde{ \map C}_{n,m} ]  \right|  \leq     \frac{ 2k d^2} {m  P_{\rm succ}    \left(\psi_x^{\otimes n}\right)   } \qquad \forall x\in\sf X \, .
\end{align*}  
In conclusion, as long as the probability of success $P_{\rm succ} (\psi_x^{\otimes n}) $ is lower bounded by a finite value independent of $m$, the $k$-copy fidelities of the cloning processes $\map C_{n,m}$ and $\widetilde{\map C}_{n,m}$.  Since the bound holds for arbitrary quantum operations, in particular it holds for the quantum operation describing the optimal cloner with given probability of success.  

It is immediate to extend the derivation to the Bayesian scenario where the input state $|\psi_x\rangle^{\otimes n}$ is given with probability $p_x$ and one considers the  average fidelity and average success probability. Indeed, the average $k$-copy fidelity is given by  
\begin{align*}  
F_{k} [\map C_{n,m}] &=    \frac{ O_{k}  [\map C_{n,m}]  }{P  [\map C_{n,m}]}  \\
\nonumber   O_{k}  [\map C_{n,m}]  & : =      \sum_x  p_x \,   O_{k,x}  [\map C_{n,m}] \\
\nonumber P  [\map C_{n,m}]  & : =   \sum_x  p_x  \,   P_{x}   [\map C_{n,m}]         
\end{align*} 
and one has the bound
\begin{align*}   
F_{k} [\map C_{n,m}]  &  \le       \frac{ 2k d^2} {m  P_{\rm succ}     }  \, ,
\end{align*} 
$P_{\rm succ}$ being the average success probability.  Again, for every fixed value of the success probability, the fidelity of the optimal cloner is achieved by a PM protocol. 

\section{Lower bound on the average probability of success}

We now show that for every fixed $n$, the probability  of success of the optimal cloner is lower bounded by a finite value.  Precisely, we prove the following
\begin{lemma}
The quantum operation $\map C_{n,m}^*$ corresponding to the  $n$-to-$m$ cloner that maximizes the fidelity in Eq. (\ref{kfid})  can be chosen without loss of generality to have success probability $P_{\rm succ}^*$ equal to $1/\| \tau^{-1} \|_{\infty} $, where $\tau$ is the average input state $\tau : =  \sum_{x}  p_x     \varphi_x^{\otimes n}$ and $\|  \tau^{-1} \|_{\infty}$ is the  maximum eigenvalue of $\tau^{-1}$. 
\end{lemma}    

{\bf Proof.}   For a generic quantum operation  $\map C_{n,m}$, the $k$-copy fidelity can be expressed in terms of its Choi operator $C_{n,m}$  as  
\begin{align*}
F_{k}[\map C_{n,m}] & =  \frac{{\tr}  [  \Omega  \,   C_{n,m}]  }{ {\tr} [ (I^{\otimes m} \otimes \bar\tau)  C_{n,m} ] } \\
 \Omega    &:  =\frac{1}{{\binom m k}}    \sum_{\sf S}  \sum_{x}   \,   p_x  \,  \left(  \psi_x^{\otimes k}\right)_{\sf S}  \otimes   \bar \psi_x^{\otimes n}  \, ,      
 \end{align*}  where     the outer summation runs over all $k$-element subsets  $\sf S$ of the output Hilbert spaces, $  \left(  \psi_x^{\otimes k}\right)_{\sf S}$ denotes the operator $\psi_x^{\otimes k}$ acting on the Hilbert spaces in the set $\sf S$, and  $\bar \tau$  ($\bar \psi_x$) is the complex conjugate of $\tau$  ($\psi_x$).
 Following the arguments of \cite{fiuraclonS,xieS}, one can easily show that   
the maximum fidelity over all quantum operations is given by 
\begin{align*} F^*_{k}  =  \left\| \left(I^{\otimes m} \otimes \bar \tau^{-\frac 12}\right )  \Omega   \left(I^{\otimes m} \otimes \bar \tau^{-\frac 12}\right )\right\|_{\infty}.
\end{align*} 
The maximum is achieved by choosing a Choi operator $C_{n,m}^* $ of the form 
\begin{align*}
C_{n,m}^*    =   \gamma~   \left(I^{\otimes m} \otimes \bar \tau^{-\frac 12}\right )  \ket{\Psi} \bra{  \Psi}   \left(I^{\otimes m} \otimes \bar \tau^{-\frac 12}\right ) 
\end{align*} 
where $\gamma\ge 0$ is a proportionality constant and $\ket{\Psi}$ is the eigenvector of $ \left(I^{\otimes m} \otimes \bar \tau^{-\frac 12}\right )   \Omega   \left(I^{\otimes m} \otimes \bar \tau^{-\frac 12}\right )$ with maximum eigenvalue.   
With this choice, the success probability $P[\map C_{n,m}^*]$ is given by 
\begin{align*}  P[\map C_{n,m}^*]  & =  {\tr}  [  C_{n,m}^*  \left(I^{\otimes m}  \otimes  \bar \tau \right)]   \\
 &=  \gamma \, .
 \end{align*}   
 We now  show that $\gamma$ can be always chosen to be larger than $1/ \|  \tau^{-1}  \|_{\infty}$.   To this purpose, note that the only constraint on $\gamma$ is that the quantum operation $\map C_{n,m}^* $ must be trace non-increasing. Now, for a generic state $\rho$ one has  
\begin{align*}
\tr[\map C_{n,m}^*  (\rho)]  &  =  \gamma ~   {\tr}[  C_{n,m}^*  \left(I^{\otimes M}  \otimes \bar \rho \right)  ] \\
  &  =  \gamma~  \bra{ \Psi}     \left(I^{\otimes M}   \otimes  \bar \tau^{-\frac 12}\bar    \rho  \bar \tau^{-\frac 12} \right)   \ket {\Psi}\\
  & \le \gamma~ \|       \bar \tau^{-\frac 12}\bar    \rho  \bar \tau^{-\frac 12}   \|_{\infty}\\
  & \le \gamma ~  \|  \tau^{-1}\|_{\infty}  \|  \rho\|_{\infty}\\
  & \le \gamma~     \|  \tau^{-1}\|_{\infty}.
\end{align*}       
 Hence, the choice $\gamma =1/  \|  \tau^{-1}\|_{\infty} $ leads to a legitimate (trace non-increasing) quantum operation.  $\blacksquare$ 


\begin{thebibliography}{99}
\bibitem{nota} The difference may  be larger at the ends of the spectrum, where the density of points is generally more rarefied.
\bibitem{nota2} This expression is also correct for small $n$. For large $n$, we can replace $|\widetilde{\mathscr P}_n|\Delta V^*$ by $V_n$, the volume of the minimal polytope that contains the lattice $\widetilde{\mathscr P}_n$.
\end{thebibliography}
\end{document}